\documentclass[twocolumn]{aastex631}

\usepackage{hyperref}

\shortauthors{Hegde et al.}

\received{March 26, 2025}
\revised{May 13, 2025}
\accepted{May 27, 2025}

\submitjournal{The Astrophysical Journal}

\begin{document}

\title{Ensemble Modeling of the Solar Wind Flow with Boundary Conditions Governed by Synchronic Photospheric Magnetograms. I. Multi-point Validation in the Inner Heliosphere}

\correspondingauthor{Dinesha Hegde}
\email{dinesha.hegde@uah.edu}

\author[0000-0002-6055-1192]{Dinesha V. Hegde}
\affiliation{Department of Space Science, The University of Alabama in Huntsville, AL
35805, USA}
\affiliation{Center for Space Plasma and Aeronomic Research, The University of Alabama in Huntsville, AL 35805, USA}

\author[0000-0003-0764-9569]{Tae K. Kim}
\affiliation{Center for Space Plasma and Aeronomic Research, The University of Alabama in Huntsville, AL 35805, USA}

\author[0000-0002-6409-2392]{Nikolai V. Pogorelov}
\affiliation{Department of Space Science, The University of Alabama in Huntsville, AL
35805, USA}
\affiliation{Center for Space Plasma and Aeronomic Research, The University of Alabama in Huntsville, AL 35805, USA}

\author[0000-0001-9498-460X]{Shaela I. Jones}
\affiliation{NASA Goddard Space Flight Center, Greenbelt, MD 20771, USA}
\affiliation{Catholic University of America, Washington, DC 20064, USA}

\author[0000-0001-9326-3448]{Charles N. Arge}
\affiliation{NASA Goddard Space Flight Center, Greenbelt, MD 20771, USA}

\begin{abstract}

The solar wind (SW) is a vital component of space weather, providing a background for solar transients such as coronal mass ejections, stream interaction regions, and energetic particles propagating toward Earth. Accurate prediction of space weather events requires a precise description and thorough understanding of physical processes occurring in the ambient SW plasma. Ensemble simulations of the three-dimensional SW flow are performed using an empirically-driven magnetohydrodynamic heliosphere model implemented in the Multi-Scale Fluid-Kinetic Simulation Suite (MS-FLUKSS). The effect of uncertainties in the photospheric boundary conditions on the simulation outcome is investigated. The results are in good overall agreement with the observations from the Parker Solar Probe, Solar Orbiter, Solar Terrestrial Relations Observatory, and OMNI data at Earth, specifically during 2020-2021. This makes it possible to shed more light on the properties of the SW propagating through the heliosphere and perspectives for improving space weather forecasts.

\end{abstract}

\keywords{Space Weather; Heliosphere; Magnetohydrodynamic simulations; Solar Wind; Interplanetary magnetic field; Parker Solar Probe; Solar Orbiter}

\section{Introduction} \label{sec:intro}

The solar wind (SW) is a stream of highly ionized plasma from the solar corona that carries the frozen-in magnetic field and extends throughout the heliosphere, as was first described by \citep{Parker:1958}. It plays a fundamental role in shaping space weather (SWx). Interaction of strong, long-duration, southward-oriented magnetic fields with the Earth's magnetic field induces geomagnetic storms \citep{Gonzalezetal:1994}. Transient solar events, such as coronal mass ejections (CMEs), are accompanied by large-scale expulsions of plasma from the solar corona, which propagate through the ambient SW \citep{Caseetal:2008} and can trigger intense geomagnetic storms \citep{Goslingetal:1990, Gosling:1993, Zhangetal:2007}. Moreover, the ambient SW is characterized by prominent large-scale structures, including high-speed streams (HSSs) and associated stream interaction regions (SIRs) formed due to the interaction of fast and slow SW. They are crucial drivers of recurrent geomagnetic storms \citep{Richardsonetal:2002, Tsurutanietal:2006}. Indeed, SIR-driven geomagnetic storms are distinctly different from the CME-driven ones.\citep{BorovskyandDenton:2006}.  Thus, an accurate description of the ambient SW, encompassing its large-scale structures and dynamics, is essential for SWx forecasting as well as for fundamental research.

Global, three-dimensional (3D) magnetohydrodynamic (MHD) models of the SW are now at the forefront of SWx research and operations. These models provide us with a detailed understanding of the large-scale, spatial, and temporal properties of the SW by combining the dynamics of the magnetic field and plasma \citep{Gombosietal:2018, Feng:2020}. Additionally, they are useful for the interpretation of coordinated remote and in situ observations, as they can approximately link observations to their sources at the surface of the Sun \citep[see, e.g.,][]{Odstrciletal:2020, Badmanetal:2023}.  

Over the past decades, several 3D MHD models have been developed \citep{Hanetal:1988, Linkeretal:1990, Usmanov:1993, Mikicetal:1999, Linkeretal:1999, Odstrcil:2003, Hayashi:2005, Nakamizoetal:2009, Fengetal:2010, detman2011,Rileyetal:2011, VanderHolstetal:2014, Shiotaetal:2014, Pogorelovetal:2009, Pogorelovetal:2014, PomoerllandPoedts:2018, Wuetal:2020} to simulate the SW. Owing to the distinct dynamical processes and computational complexities involved, the modeling space is generally divided into two regions: the solar corona, which extends from the solar surface to a region slightly above the Alfv\'en surface (typically at $20$–$30,\mathrm{R_{\odot}}$), and the inner heliosphere (IHS), which spans from the outer boundary of the solar corona to Earth and beyond (upto few $au$).
Depending on the modeling strategy, these models can be broadly categorized into fully physics-based models, which simulate both domains using a single or separate MHD frameworks, and mixed models, which employ relatively simple, semi-empirical and/or kinematic techniques for coronal modeling, while treating the IHS using MHD model \citep{WuandDryere:2015, Feng:2020}. Examples of such semi-empirical coronal models include the Wang--Sheeley--Arge (WSA)  \citep{ArgeandPizzo:2000, Argeetal:2003} and Hakamada--Akasofu--Fry (HAF) \citep{HakamadaandAkasofu:1981, Fryetal:2001} models, and the Distance from the Coronal Hole Boundary (DCHB) method \citep{Rileyetal:2001}.

Notably, the mixed SW models, e.g., WSA-ENLIL \citep{Odstrciletal:2005,Pizzoetal:2011}, EUHFORIA \citep{PomoerllandPoedts:2018} have proved to be rather effective for operational SWx forecasts due to their computational efficiency. However, the ability to make accurate predictions of SW remains far from the desired level. The predictive capabilities of various global SW models have been extensively evaluated in the literature \citep{MacNeiceetal:2018}. For example, \citet{Jianetal:2015} conducted a comprehensive comparison of SW models implemented at the NASA Community Coordinated Modeling Center (CCMC) and demonstrated that, while each model exhibited its own strengths, none of them consistently outperformed the others in reproducing all aspects of SW conditions at Earth. Thus, a systematic investigation of the sources leading to inaccuracies in all those models is a critical first step towards understanding their true capabilities and limitations.

Most global SW models primarily rely upon the solar photospheric magnetic field distributions as observational input to establish inner-boundary conditions. However, continuous observations of the magnetic field are available only for the near-side of the Sun. They cover only a part of the solar surface, leaving the far side and polar regions unobserved. To compensate for this observational limitation, early modeling approaches adopted a steady-state (in a corotating frame) assumption and constructed traditional synoptic maps by combining the ground or space based line-of-sight (LOS) full-disk magnetograms along the central meridian over a full solar rotation period, also called the Carrington rotation (CR), i.e., $\approx$ 27 days \citep{HarveyandWorden:1998, Liuetal:2017}. However, these maps do not capture a number of flux evolution processes, e.g., differential rotation and meridional flows, occurring on a much shorter time scale, which can significantly affect the large-scale coronal and heliospheric magnetic structure \citep{Argeetal:2010}.

To address those shortcomings, various surface flux transport (SFT) models \citep{SchrijverandDeRosa:2003, Argeetal:2010, Hickmannetal:2015, UptonandHathaway:2014,Pogorelovetal:2024,Caplanetal:2025} have been developed to simulate the continuous evolution of the photospheric magnetic field across the entire solar surface. These models produce instantaneous global magnetic field maps, known as synchronic maps, by assimilating available near-side observations, thereby providing more realistic and time-dependent boundary conditions for SW simulations. 

However, the SFT models are not free from uncertainties themselves. They arise from the underlying assumptions, parameter choices, and assimilation of observational data. Since many of the physical processes governing the evolution of the photospheric magnetic field remain poorly understood, and due to the lack of instantaneous observations of the entire solar surface, reducing uncertainties in the SFT model remains a significant challenge. Consequently, SFT modelers typically generate multiple realizations (an ensemble) of synchronic maps to represent the plausible range of variability in a model as a part of uncertainty quantification. 

Given the inherent uncertainties associated with SFT models, it is essential to characterize their impact on SW simulations. The most viable approach to address this is to take into account the uncertainties through the implementation of ensemble boundary conditions for the coronal and inner-heliospheric models \citep[e.g.,][]{Poduvaletal:2020}. In the context of uncertainty quantification, ensemble modeling broadly refers to the practice of running multiple simulations with varied inputs that reflect known or estimated inherent uncertainties \citep{Wilks:2019}. Within this framework, ensemble modeling enables assessment of the model sensitivity to those uncertainties and provides means to quantify confidence in the resulting predictions. Beyond offering a spectrum of plausible outcomes, it allows for probabilistic interpretation, which offers a clear advantage over deterministic forecasts and ultimately improves the robustness of space weather predictions under both observational and physical constraints.

Moreover, to improve the reliability and accuracy of the models, their regular performance evaluation by validating solutions with observations is essential. Generally, most global 3D MHD models of the SW in the IHS are validated and optimized using near-earth in situ SW plasma and magnetic field observations from spacecraft, such as the Advanced Composition Explorer (ACE) \citep{Stoneetal:1998} and Wind \citep{Acunaetal:1995}, which orbit at L1 (the first Lagrange point of the Sun-Earth system) \citep{Leeetal:2009, Gressletal:2014, Jianetal:2015, Reissetal:2016}.  

Meanwhile, recently launched heliospheric missions such as the Parker Solar Probe (PSP) \citep{Foxetal:2016} and Solar Orbiter (SolO) \citep{Mulleretal:2013, Mulleretal:2020} are providing an unprecedented set of in situ SW measurements across a wide range of radial, latitudinal, and longitudinal positions within the IHS. Furthermore, this dataset is complemented by observations from the Solar Terrestrial Relations Observatory (STEREO-A) \citep{Kaiseretal:2008}. Consequently, these diverse, multi-point observations present a unique opportunity to enhance the validation and refinement of data-driven numerical models, offering a more comprehensive assessment of SW dynamics across the IHS.

Over the years, many multi-spacecraft validation studies of inner heliospheric 3D MHD models of the SW have been conducted using prior \citep{Merkinetal:2016, Wangetal:2020, Lietal:2020}, as well as recent \citep{Rileyetal:2021, Zhangetal:2023, Knizhniketal:2024} space missions. Such studies have demonstrated the feasibility and value of multi-spacecraft observations in evaluating the accuracy of the applied model.

In this work, we perform simulations of the ambient SW flow in the IHS utilizing the MHD model implemented in the Multi-Scale Fluid-Kinetic Simulation Suite (MS-FLUKSS) \citep{Pogorelovetal:2014} coupled with the semi-empirical WSA coronal model \citep{ArgeandPizzo:2000, Argeetal:2003, Argeetal:2004}.  To perform WSA simulations, we use synchronic maps from the Air Force Data  Assimilative Photospheric Flux Transport (ADAPT) model \citep{Argeetal:2013, Hickmannetal:2015} constrained by the photospheric magnetic field observations from the Helioseismic and Magnetic Imager (HMI) \citep{Scherreretal:2012} onboard the Solar Dynamics Observatory (SDO) \citep{Pesneletal:2012}.

We employ an ensemble modeling technique to investigate the uncertainty in the SW simulation arising from the photospheric boundary conditions intrinsic to the ADAPT model. While performing these simulations, we compare them with the multi-point in situ observations along the PSP, Earth, SolO, and STEREO-A trajectories during the ascending phase of solar cycle 25. We qualitatively evaluate the performance of our model in reproducing the large-scale dynamics of the SW.  A simultaneous, quantitative uncertainty and performance analysis at multiple locations is not a trivial matter, so we reserve it for the second part of this paper.

This paper is structured as follows. Section \ref{sec:SWmodel} describes the SW model, including the inner boundary conditions, uncertainty quantification, and ensemble modeling. Further, section \label{subsec:Obs&Valid} details the in situ observations from multiple spacecraft and their configuration during the selected validation period. Section \ref{sec:Results} presents the simulation results, comparing the model solutions with in situ data for each validation period separately. Finally, section \ref{sec:DandC} discusses our findings and provides the conclusions.

\section{Solar Wind Model} \label{sec:SWmodel}

\subsection{Inner Heliospheric Model: MS-FLUKSS} \label{subsec:IHmodel}

MS-FLUKSS is a collection of highly parallelized numerical modules designed for modeling the flow of partially ionized plasma in the 3D global heliosphere and its interaction with the local interstellar medium across multiple scales on adaptive grids. The block structure of MS-FLUKKS is outlined in \cite{Pogorelovetal:2014}. Its new developments are described by \cite{pogorelov2016, pogorelov2017a,pogorelov2017b,pogorelov2021,Fraternale_2023} and \cite{Bera2023}.

MS-FLUKSS employs ideal MHD equations for ion modeling, while neutrals are addressed either kinetically or through a multi-fluid approach. The suite is designed to support multiple coordinate systems and features an Adaptive Mesh Refinement (AMR) capability, which strategically enhances mesh resolution in targeted areas and is built upon the Chombo architecture \citep{Colellaetal:2007}. It can accommodate 3D time-dependent inner boundary conditions derived either through SW observations or from different models \citep{borov12,pogorelov2013a, Kimetal:2016, Kimetal:2020}. Furthermore, MS-FLUKSS is equipped with multiple CME models \citep{Pogorelovetal:2017, Singhetal:2020a, Singhetal:2020b, Singhetal:2022, Singhetal:2023}, designed to simulate CME propagation within the SW background of the IHS. In this work, we focus solely on modeling the ambient SW along the trajectories of multiple spacecraft in the IHS.

We model the ambient supersonic SW by solving the set of single fluid ideal MHD equations, which represent the conservation of mass, momentum, energy, and magnetic flux:

\begin{equation}
    \frac{\partial \rho}{\partial t} + \nabla \cdot [\rho \mathbf{v}] = 0
    \label{eq:continuity}
\end{equation}

\begin{equation}
    \frac{\partial \rho \mathbf{v}}{\partial t} + \nabla \cdot \left[\rho \mathbf{v} \mathbf{v} + p_{0} \hat{\mathbf{I}} - \frac{\mathbf{B} \mathbf{B}}{4\pi}\right] = 0
    \label{eq:momentum}
\end{equation}

\begin{equation}
    \frac{\partial e}{\partial t} + \nabla \cdot \left[(e + p_0 ) \mathbf{v} -  \frac{\mathbf{B} \left(\mathbf{v} \cdot \mathbf{B}\right)}{4\pi}\right] = 0
    \label{eq:energy}
\end{equation}

\begin{equation}
    \frac{\partial \mathbf{B}}{\partial t} + \nabla \cdot [\mathbf{v} \mathbf{B} - \mathbf{B} \mathbf{v}] = 0
    \label{eq:induction}
\end{equation}

Here, $\rho$ represents the mass density, while $\mathbf{v}$ and $\mathbf{B}$ denote the velocity and magnetic field vectors, respectively. Furthermore, \(p_0 = p + B^2/8\pi\) denotes the total pressure, which is the sum of thermal pressure (\(p\)) and magnetic pressure. Additionally, \(e = p/(\gamma - 1) + \rho v^2/2 + B^2/8\pi\) is the total energy density, with \(\gamma\) representing the adiabatic index. Lastly, \(\hat{\mathbf{I}}\) is the identity tensor. We assume \(\gamma = 1.5\) to represent the polytropic relation between density and pressure in the IHS \citep[e.g.,][]{Tottenetal:1995, Rileyetal:2001, Merkinetal:2016, Kimetal:2020}.

The system of hyperbolic partial differential equations (Eq.\ref{eq:continuity}-\ref{eq:induction}) is solved in a spherical coordinate system employing a Godunov-type, finite-volume numerical scheme, which guarantees the second order of accuracy in both space and time. Numerical fluxes across computational cell boundaries are calculated using a Roe-type MHD Riemann problem solver \citep{kulikovskii:2001}. The monotonized central difference slope limiters are utilized to maintain the total variation diminishing property in the solution. Furthermore, the Hancock method, known for its second-order accuracy, is applied for time integration. To preserve the $\nabla \cdot \mathbf{B} = 0$ condition in numerical simulations, we adopt the 8-wave divergence cleaning approach, modifying the ideal MHD equations with artificial source terms proportional to $\nabla \cdot \mathbf{B}$ in momentum, energy, and induction equations \citep{Powelletal:1999}. Additionally, to track and resolve the heliospheric current sheet (HCS) within our simulated heliospheric environment, we employ the level set method by solving an additional advection equation, $s_t+\mathbf{v} \cdot \nabla s = 0$, where $s$ denotes the position of the surface, to accurately represent the passive advection of the HCS by the SW flow \citep{Borovikovetal:2011}.

We choose a Sun-centered spherical coordinate system, with the $Z$ axis aligned with the northward solar rotation axis.  The $X$ axis lies in the plane formed by the velocity vector of the undisturbed local interstellar medium (LISM) and the $Z$ axis. That is, the $X$ axis is rotated $X$ axis of the heliographic inertial (HGI) system \citep{Burlaga:1984} about the $Z$ axis by approximately $178.98^\circ$ counterclockwise, ensuring that the Y-component of the LISM is zero. The $Y$ basis vector completes the right-handed orthogonal system \citep{Kim:2014thesis}.

We solve our equations on a spherical grid consisting of $200 \times 256 \times 128$ cells in the radial ($R$), azimuthal ($\phi$), and latitudinal ($\theta$) directions. The grid spans radially from the inner boundary at $R = 0.1,\mathrm{au}$ ($21.5,\mathrm{R_{\odot}}$) to the outer boundary at $1.1,\mathrm{au}$, employing a non-uniform exponential distribution that increases cell size in the outward direction. Specifically, $\Delta r = 0.0029\,\mathrm{au}$ ($0.63\,\mathrm{R_{\odot}}$) and $\Delta r = 0.0080\,\mathrm{au}$ ($1.71\,\mathrm{R_{\odot}}$) at the innermost and outermost cells, respectively. The grid is also non-uniform in the latitudinal direction, i.e., it is geometrically stretched near the poles (in the regions $\pm 14^\circ$ from the $z$-axis), featuring $\Delta\theta = 2.92^\circ$ in the cell adjacent to the polar axis ($\theta =0$ and $\theta =\pi$) and switching to uniform spacing of $\Delta\theta = 1.31^\circ$ in the remaining region. The grid is uniform in the $\phi$ -direction.

 \begin{figure*}[ht!]
     \centering
     \includegraphics[width=\textwidth]{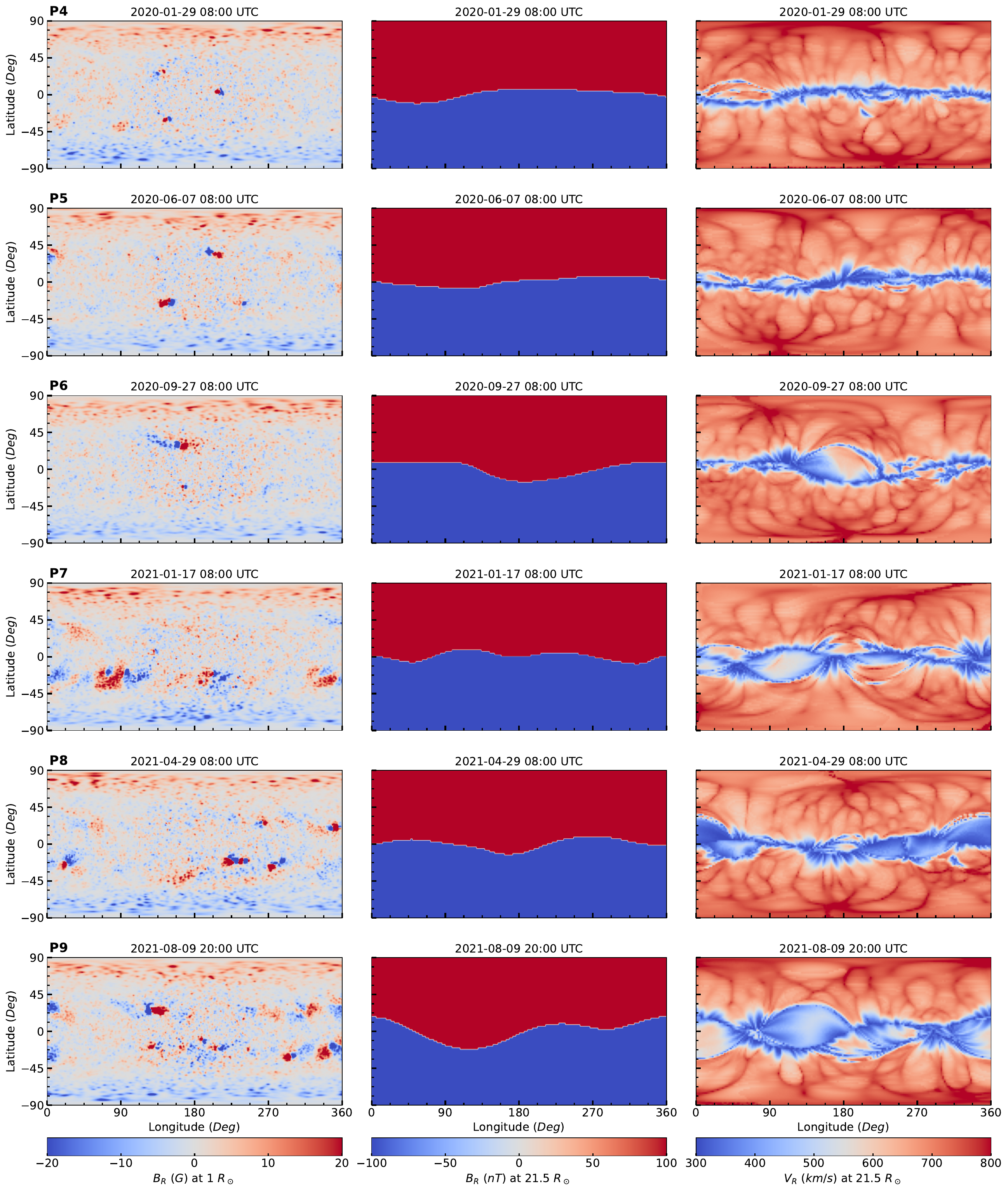}
    \caption{Example boundary maps used in our simulation correspond to the sixth to ninth PSP perihelion (P4-P9, shown from top to bottom panels) for the zeroth realization. The X and Y axes represent the solar latitudes and longitudes within the central meridian-centered heliographic frame, where Earth is always located at $180^\circ$ longitude. (left panels) The ADAPT-HMI radial magnetic field map at the solar photosphere ($1 \, R_{\odot}$), with color saturation at $\pm 20\, \mathrm{G}$. The WSA radial magnetic field ($B_{R}$) in $\mathrm{nT}$(center panels). The WSA radial velocity component ($V_{R}$) in $\ \mathrm{km}\,\mathrm{s}^{-1}$ (right panels) at $21.5 \, R_{\odot}$}
     \label{fig:bc}
 \end{figure*}

\subsection{Coronal Model: WSA}

The WSA model combines the Potential Field Source Surface (PFSS) \citep{Schattenetal:1969, AltschulerandNewkirk:1969} and Schatten’s Current Sheet (SCS) \citep{Schatten:1972} models for reconstructing the coronal magnetic field. The PFSS model assumes a current-free solar atmosphere and extrapolates the radial magnetic field from the solar photosphere to its outer boundary, the source surface, typically set at 2.5 $R_{\odot}$, solving the Laplace equation. The SCS model further extrapolates the radial magnetic field from this source surface up to the inner boundary of the IHS model by again solving the Laplace equation. This process involves first adjusting the inward magnetic field lines outward at the source surface. The final step reverses these magnetic field lines back to their original polarity, creating current sheets at regions with opposing magnetic fields. 

Following the reconstruction of the coronal magnetic field, field lines are traced from the outer boundary back to the innermost boundary, the photosphere. A map of magnetic flux tube expansion factors ($f_s$) between the photosphere and the source surface is then computed \citep{WangandSheeley:1992}. Furthermore, the coronal hole map is generated using the traced open and closed field footpoints, and the minimum angular separation ($\theta_b$) of open field footpoints from the nearest coronal hole boundary is determined on the photosphere. Finally, the radial velocity component ($V_{R}$) at the outer boundary is estimated using $f_s$ and $\theta_b$ \citep{Argeetal:2003, Argeetal:2004}, as outlined in Equation \ref{eq:wsa}.

\begin{equation}
V_R = V_0 + \frac{V_1 \left(\beta - \gamma \exp{\left[(\frac{{\theta_b}}{w})^{\delta}\right]}\right)^{3}} {(1 + {f_{s}})^{\alpha}}
\label{eq:wsa}
\end{equation}
$V_0 = 285\ \mathrm{km}\,\mathrm{s}^{-1}$, $V_1 = 625\ \mathrm{km}\,\mathrm{s}^{-1}$, $\alpha = 1/4.5$, $\beta = 1$, $w = 2$, $\gamma = 0.8$, $\delta = 2$. \\
In addition,
\begin{equation}
f_s = \left(\frac{R_{ph}}{R_{ss}}\right)^2 \frac{B_{ph}}{B_{ss}}
\label{eq:expfactor}
\end{equation}
where $R_{ph} = 1 R_{\odot}$, $R_{ss} = 2.5 R_{\odot}$, and $B_{ph}$ and $B_{ss}$ are the radial magnetic field strengths at the solar surface and source surface. 

\subsection{Surface Flux Transport Model: ADAPT}

The ADAPT model was built upon an SFT model originally introduced by \cite{WordenandHarvey:2000}, incorporating data from LOS or vector magnetograms \citep{Hickmannetal:2015}. ADAPT comprehensively simulates the distribution of the solar magnetic field, including such areas as the Sun's far side and its poles, by incorporating various aspects of magnetic flux transport. These aspects include differential rotation, meridional flow, supergranular diffusion, and random flux emergence \citep{Argeetal:2010, Argeetal:2011, Argeetal:2013}. Notably, employing an ensemble least squares technique, the ADAPT model can assimilate magnetic field observations from instruments like National Solar Observatory (NSO)/Global Oscillation Network Group (GONG) \citep{Hill:2018}, NSO/Synoptic Optical Longterm Investigations of the Sun (SOLIS)/Vector Spectromagnetograph (VSM) \citep{Kelleretal:2008}, and the SDO/HMI. To quantify the observational and model uncertainties related to the supergranular flows, ADAPT generates $12$ equally possible synchronic maps as an ensemble, each varying the supergranular distributions statistically in the far-side and polar regions on the solar surface, which lacks regular photospheric magnetic field observations \citep{Hickmannetal:2015}. 

\subsection{MHD Boundary Conditions}

The outer boundary of the MHD model is set to encompass the near-Earth environment and the orbits of spacecraft considered for validation in this study. No boundary conditions are necessary at the outer boundary surface, so we apply a second-order extrapolation of the SW properties into the layer of ghost cells surrounding the outer boundary.

We set our inner boundary of the MHD model just above the critical spherical surface where the radial velocity of the SW exceeds the fast magnetosonic speed—in other words, where the SW becomes superfast magnetosonic. Recent observations from PSP indicate that the Sun's Alfv\'en critical surface is located between 16 and 20 \(R_\odot\) \citep{Kasperetal:2021}.
As PSP descends below $0.1\,\mathrm{au}$, reaching $0.09 \,\mathrm{au}$ ($20.3\, \mathrm{R_{\odot}}$) during perihelion passes 6 and 7 (P6 \& P7), we adjust the inner boundary to $19\, \mathrm{R_{\odot}}$. However, for P8 and P9, when PSP reaches a heliocentric distance of $0.074\,\mathrm{au}$ ($15.9\, \mathrm{R_{\odot}}$), we retain the default inner boundary at $21.5\, \mathrm{R_{\odot}}$, as we cannot extend it below the critical surface. 

We use the results from the WSA model version-6 (WSA-v6.0)\footnote{\url{https://ccmc.gsfc.nasa.gov/models/WSA~6/}} driven by the ADAPT maps constrained by the SDO/HMI magnetograms to establish the inner boundary conditions for our MHD model. We keep the same set of WSA free parameters, as shown in Eq.~\ref{eq:wsa}, through all our simulations.

The original WSA maps of $2^{\circ} \times 2^{\circ}$ spatial resolution and 12-hour cadence are interpolated onto the MS-FLUKSS grid in both space and time. While incorporating the WSA boundary conditions into MS-FLUKSS, we perform a longitudinal shift or rotation of a certain degree (e.g., 10$^\circ$ at 21.5 $R_{\odot}$) to account for the solar rotational effect during the travel time of solar wind from the Sun's surface to the IHS inner boundary \citep[e.g.,][]{MacNeiceetal:2011}. 

Furthermore, we scale the WSA radial magnetic field by a factor of $2$ before mapping it onto the MS-FLUKSS inner boundary to compensate for the systematic underestimation of magnetic field strengths at 1 $au$ \cite[e.g.,][]{Linkeretal:2016}, known as the ``open flux'' problem \citep{Linkeretal:2017}. Additionally, we reduce the WSA velocities uniformly by $75\ \mathrm{km}\,\mathrm{s}^{-1}$ to account for the difference in SW acceleration between the WSA kinematic and MHD models \citep[e.g.,][]{MacNeiceetal:2011, Kimetal:2014}. We assume that the nonradial components of the SW velocity are zero.

As a result of solar rotation, the heliospheric field in our inertial coordinate system acquires an azimuthal component, \(B_{\phi}\), absent in the WSA model. The calculation of \(B_{\phi}\) is given by \(B_{\phi} = - \Omega R B_{R} \sin\theta /V_R\), where \(\Omega\) represents the solar angular rotation rate, \(R\) the radial distance from the Sun, \(B_R\) the radial component of the magnetic field, \(\theta\) represents the latitude and \(V_R\) the radial SW velocity derived from WSA. Furthermore, we adjust the \(B_R\) to preserve the original open magnetic flux. Finally, the latitudinal component of the magnetic field, \(B_{\theta} \), is assumed to be zero \citep{Parker:1963}. 

We establish the SW density and temperature at the inner boundary by assuming momentum flux and thermal pressure equilibrium, as given by ad hoc prescriptions $N  V_R^{2} = N_\mathrm{min}  V_\mathrm{fast}^{2}$ and $N T = N_\mathrm{min} T_\mathrm{max}$, where $N$ is the SW density in cm$^{-3}$, $V_R$ is the SW radial velocity component in $\mathrm{km}\, \mathrm{s}^{-1}$, $T$ is the temperature in $10^6$~K, $N_\mathrm{min}= 200$ {cm}$^{-3}$, 
$T_\mathrm{max} =2 \times 10^{6}$~K,  $V_\mathrm{fast} = 700\ \mathrm{km}\, \mathrm{s}^{-1}$, and the minimum SW velocity is set 
to 200 $\mathrm{km}\, \mathrm{s}^{-1}$\citep[e.g.,][]{Jianetal:2016, Kimetal:2020}.

Figure \ref{fig:bc} presents example boundary maps derived from the ADAPT and WSA models for the fourth to ninth PSP perihelia (P4–P9, shown from top to bottom panels) for the zeroth realization. The left panels display the ADAPT-HMI radial magnetic field component, $B_R$, at the solar photosphere, while the center panels show $B_R$ from the WSA model, and the right panels present the WSA radial velocity component, $V_R$, at the inner boundary of MS-FLUKSS. These maps depict the global distribution of parameters across heliographic longitudes and latitudes within a central meridian-centered heliographic frame \footnote{\url{https://nso.edu/data/nisp-data/adapt-maps/}}, where the Sun-Earth line aligns with the $180^\circ$ heliolongitude.

In Figure \ref{fig:bc}, the central high-resolution approximately circular patch on the ADAPT $B_R$ maps represents the HMI LOS magnetogram data assimilation window. The WSA $B_R$ maps show the presence of an HCS separating the positive and negative polarity sectors and reveal its slight tilt relative to the solar equator. Notably, these maps show the smoothed global dipolar magnetic field in contrast to the highly structured photospheric magnetic field shown in ADAPT $B_R$ maps. The $V_R$ maps show the slow SW distributed near the solar equator, enveloping the HCS, and faster winds at higher latitudes. The WSA $B_R$ and $V_R$ maps exhibit the typical features associated with the near-minimum/ascending phase of the solar cycle. The blobs of slow and fast wind mixtures (e.g., at $180^\circ$ longitude in the P6 $V_R$ map) result from the interactions between the fast wind emanating from the equatorial extensions of the polar coronal holes and the equatorial slow SW. The ADAPT and WSA solutions for different perihelia during the rising phase of the solar cycle show that there is a gradual increase in the number of active regions, the tilt of the HCS relative to the solar equator, and the latitudinal extent of the slow SW region.

\subsection{Uncertainty Quantification via Ensemble Modeling}

We employ the publicly available ADAPT global maps based on SDO/HMI LOS magnetograms, which are produced every $12$ hours and feature $12$ ensemble members (or realizations) to serve as the inner boundary conditions for the WSA model. These HMI-ADAPT-WSA maps are then employed as an ensemble of inner boundary conditions for subsequent simulations in the IHS. This approach allows for rigorous quantification and propagation of uncertainties from the solar surface, through coronal modeling, and into the MHD model of the SW.

 \begin{figure*}[ht!]
     \centering
     \includegraphics[width=\textwidth]{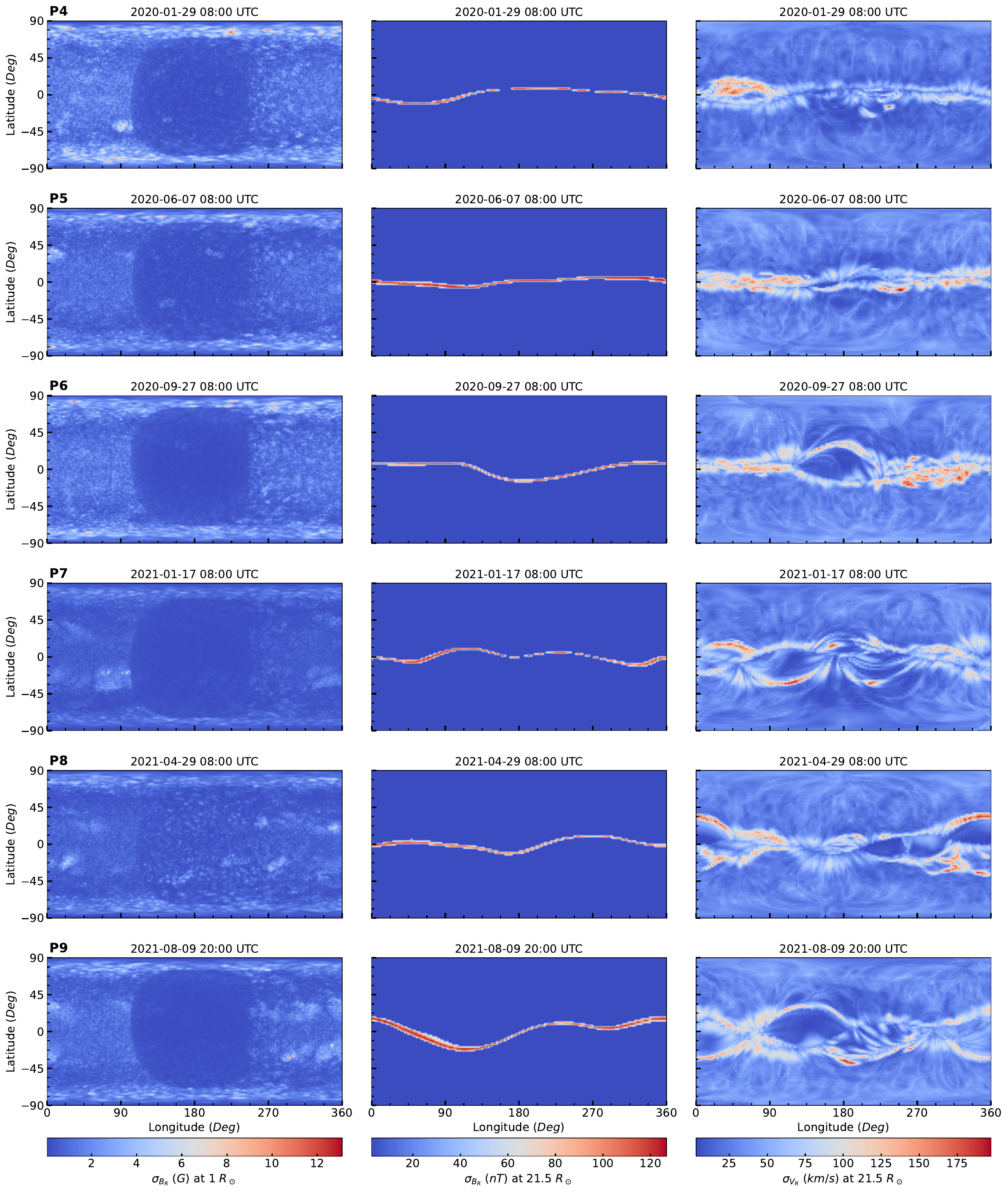}
    \caption{Standard deviation maps of the 12 realizations of the global boundary maps used in our simulation correspond to the sixth to ninth PSP perihelia (P4-P9, shown from the top to the bottom). The format of the plot is the same as in Figure \ref{fig:bc}.}
     \label{fig:bcstd}
 \end{figure*}
 
Figure \ref{fig:bcstd} presents standard deviation maps calculated from the 12 ensemble members for P4–P9 (shown from the top to the bottom) in the same format as in Figure \ref{fig:bc}. The left panels display the standard deviation of the ensemble maps of ADAPT-HMI $B_R$, at the solar photosphere, while the center and right panels show the standard deviation of the ADAPT-driven WSA coronal $B_R$ and $V_R$ ensemble maps at $21.5 \, R_{\odot}$, respectively. These maps represent the global distribution of uncertainty levels, as measured by the spread among the ensemble members, and illustrate how uncertainty in the photospheric boundary conditions propagates into the inner boundary conditions of the IHS model through the WSA coronal model.

As shown in Figure \ref{fig:bcstd}, the spread in the ADAPT-HMI $B_R$ is primarily concentrated at the polar regions (i.e., latitudes above $70^\circ$), while relatively smaller spreads exist on the far side of the Sun (i.e., longitudes below $110^\circ$ and above $250^\circ$). In contrast, the central region (centered at $180^\circ$ longitude) exhibits a distinctive blue patch with minimal spread, as this area represents the region of HMI magnetogram data assimilation common to all ADAPT ensemble members.

Regarding the coronal magnetic field, the WSA-$B_R$ standard deviation maps reveal negligible variations among the ensemble members across most regions, except near the HCS. The increase in the standard deviation near the HCS primarily arises from the changes in its position. This indicates that uncertainties in the photospheric magnetic field predominantly affect the position of the HCS, which in turn impacts the magnetic field polarity and magnitude near the HCS. Interestingly, the maps across different PSP perihelia with varying HCS tilts demonstrate that larger HCS tilts correspond to greater positional changes in the HCS.

Furthermore, the WSA-$V_R$ standard deviation maps show that the spread in the ensemble members remains relatively unchanged at higher heliographic latitudes, while larger variations are concentrated around certain low latitudes near the equator. Comparing these maps with the WSA-$V_R$ maps in Figure \ref{fig:bc}, it is evident that the largest variations occur at the boundary between the slow and fast SW. These variations arise from changes in the position of the HCS, as the changes in HCS location alter the positions of traced open magnetic field line footpoints from one polarity to the other. This, in turn, modifies the location of the coronal hole boundaries, which ultimately determines the boundaries of the fast and slow SW in the WSA framework. Notably, the standard deviation of the WSA $V_R$ realizations can reach up to $195 \, \mathrm{km}\,\mathrm{s}^{-1}$, which will propagate further into the IHS domain as inner boundary conditions.

\section{Observations and Validation Strategy} \label{subsec:Obs&Valid}

\subsection{In situ Solar Wind Data} \label{subsec:SWObs}
 
In our study, we use the in situ SW measurements along the trajectories of Earth, PSP, SolO, and STEREO-A to validate our model. We consider essential plasma parameters such as the SW velocity, number density, and temperature, along with the magnetic field, crucial for characterizing the large-scale nature of the SW. We use the so-called spacecraft RTN coordinate system, where the basis vectors are defined by the radial ($R$) direction, which aligns with the Sun-spacecraft line; the tangential ($T$) direction, corresponding to the cross product of the solar rotation axis and the R vector; and the normal ($N$) direction, which completes the right-handed orthogonal triad.

For PSP, we utilize the Maxwellian fits of particle velocity distributions from Level-3 ion data, measured by the Solar Probe Cup (SPC; \citealp{Caseetal:2020}), an instrument in the Solar Wind Electrons, Alphas, and Protons (SWEAP) suite \citep{Kasperetal:2016}, to extract the plasma properties. In instances where SPC data are not viable, particularly around perihelia, we employ partial moments of the proton distribution functions obtained from another instrument in the SWEAP suite, the Solar Probe Analyzer-Ions (SPAN-I; \citealp{Livietal:2022}). A general quality flag has been applied to proton number density and the temperature data from SPC to filter out bad-quality data. The magnetic field is obtained from the Level-2 data of the flux-gate magnetometer of the Electromagnetic Fields Investigation (FIELDS; \citealp{Baleetal:2016}) instrument onboard the PSP. All data obtained from SPC, SPAN-I, and FIELDS are given in the RTN coordinate system and are reduced to a 1-hour temporal resolution.

For SolO, our study utilizes Level-2 magnetic field data from the flux-gate magnetometer (MAG) \citep{Horburyetal:2020} and Level-2 plasma moments data from the Solar Wind Analyser (SWA) suite \citep{Owenetal:2020}. These data (given in the RTN frame) are down-sampled to an hourly time cadence by applying the necessary quality flags.

At Earth, the SW data in the RTN frame are directly retrieved with a 1-hour cadence from the OMNI database \citep{Papitashvili:2005}. This repository compiles the SW measurements from various spacecraft in halo orbits around the Sun-Earth L1 point and adjusts
them for the presence of the Earth's magnetosphere. The primary sources of these data are the ACE \citep{Stoneetal:1998} and Wind \citep{Acunaetal:1995} spacecraft. The Magnetic Field Experiment (MAG) \citep{Smithetal:1998} and the Solar Wind Electron, Proton, and Alpha Monitor (SWEPAM) \citep{MaComasetal:1998} onboard ACE collect the magnetic field and plasma data. Meanwhile, the Solar Wind Experiment (SWE) \citep{Ogilvieetal:1995} measures the plasma data, while the Magnetic Field Investigation (MFI) \citep{Leppingetal:1995} onboard WIND gathers the magnetic field data.

For STEREO-A, we use magnetic field measurements from the in situ Measurements of Particles and CME Transients (IMPACT) \citep{Luhmannetal:2008} and plasma flow properties from the PLAsma and SupraThermal Ion Composition (PLASTIC) \citep{Galvinetal:2008} instruments. Here, we directly obtain the 1-hour cadence data in the RTN frame. 

Since our goal is to validate only the background SW solutions in the IHS without large-scale transients, it was necessary to exclude periods of interplanetary CMEs (ICMEs) observed by the spacecraft. To identify the ICME intervals, we utilized a recent and comprehensive HELIO4CAST ICME catalog \citep{Moestletal:2020} of ICMEs observed by the Wind, PSP, SolO, and STEREO-A spacecraft during our validation period. 

\subsection{Validation Periods and Spacecraft Configurations}

 \begin{figure*}[t]
     \centering
     \includegraphics[width=\textwidth]{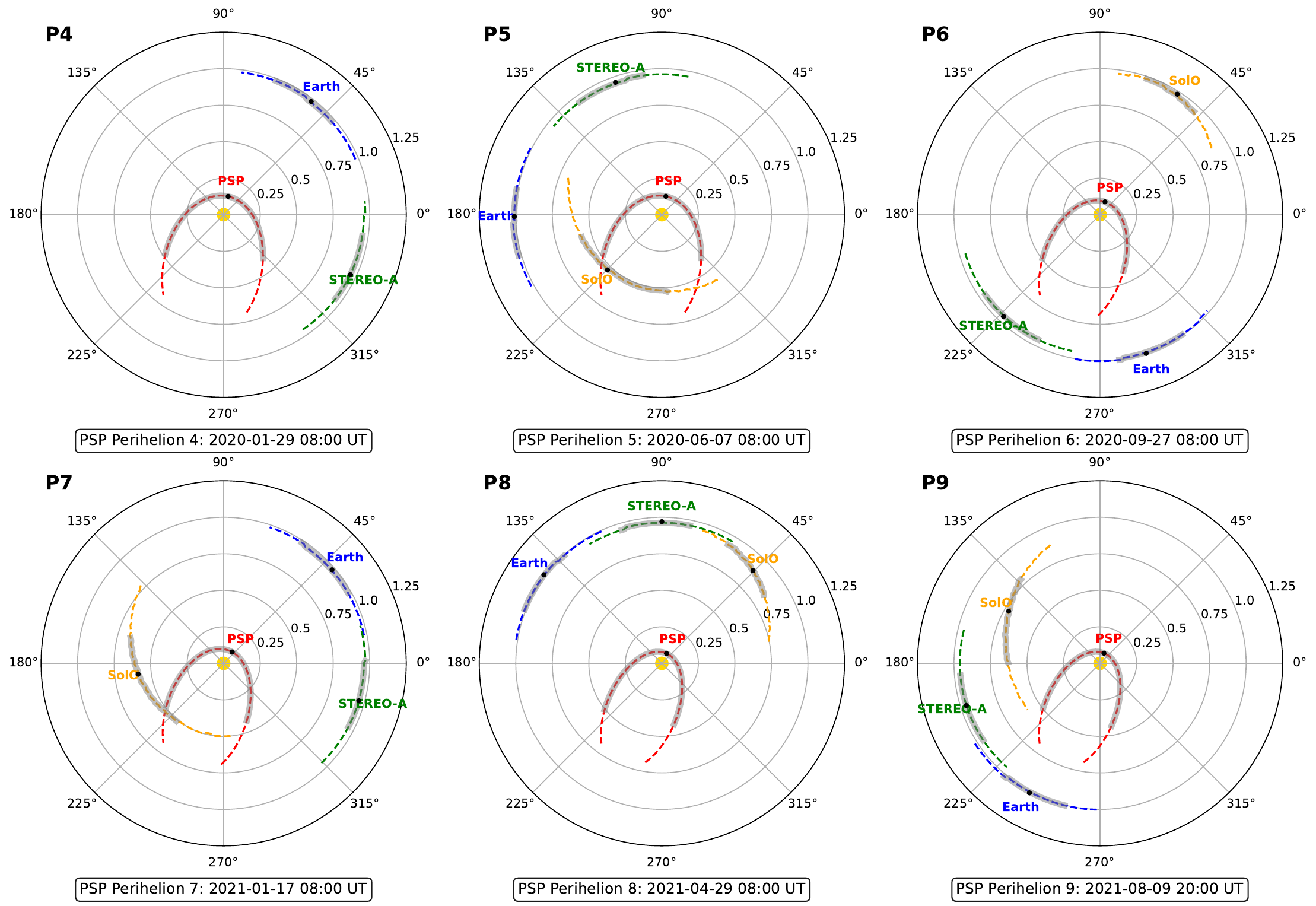}
    \caption{Polar plots showcasing the orbits of Earth (blue), STEREO-A (green), SolO (orange), and PSP (red) in the inner heliosphere during the time interval between the fourth and the ninth PSP perihelion passes (P4 to P9). The radial and polar axes indicate the heliocentric radial distances in $au$ and helio-longitudes in degrees, respectively. These positions are shown in the HGI coordinate system, projected onto the equatorial plane. The dashed curves illustrate the trajectories of each spacecraft in the range of $\pm 30$ days around the PSP perihelion, with all trajectories following a counterclockwise direction in the heliocentric orbit. The approximate spacecraft locations are highlighted with black dots and labeled accordingly for each spacecraft. The shaded segments for each trajectory indicate the time frames around the perihelion passes, specifically 12 days before and 16 days after the PSP perihelion. These intervals are the periods during which we validate our SW simulations in this study.}
     \label{fig:traj}
 \end{figure*}

We compare our simulation results in the IHS with in situ measurements along the Earth, PSP, SolO, and STEREO-A trajectories during multiple PSP solar encounters. Considering that our model results for the first three PSP perihelion passes and their corresponding full orbits were already presented in \citet{Kimetal:2020}, here we focus on the subsequent perihelion passes that occurred during 2020-2021 (P4- P9). These intervals are defined as the time frames starting 12 days before (inbound phase) and ending 16 days after (outbound phase) each PSP perihelion. This validation window was chosen based on the availability of PSP data and to ensure coverage of at least one solar rotation. Each PSP perihelion pass provides a unique spacecraft configuration, offering valuable opportunities to assess the performance of our model at different locations in the IHS.

In Figure \ref{fig:traj}, the radial and longitudinal configurations of the spacecraft trajectories for P4--P9 are shown as polar plots to provide a visual context for the validation points in the IHS. The spacecraft positions are shown in the HGI coordinate system, also known as the heliocentric inertial (HCI) system, where the $Z_{HGI}$ basis vector aligns with the Sun's rotation axis pointing northward, the $X_{HGI}$ basis vector corresponds to the line formed by the intersection of the solar equatorial plane and the ecliptic plane, known as the longitude of the ascending node. Finally, the $Y_{HGI}$ basis vector completes the right-handed orthogonal triad \citep{Burlaga:1984}. The spacecraft trajectories corresponding to the time frames of perihelion passes are shown as shaded segments, while the dashed curves represent the $\pm 30$ days around the PSP perihelion. All spacecraft follow a counterclockwise direction in the heliocentric orbit. The approximate spacecraft locations on the day of each perihelion pass are marked with black dots and labeled with different colors.

As shown in Figure \ref{fig:traj}, the longitudinal sweep of Earth, STEREO-A, and SolO was comparable throughout the validation period, whereas PSP’s sweep was noticeably larger due to its higher orbital velocity. PSP followed a similar trajectory during P4–P5, P6–P7, and P8–P9, but its relative position to Earth varied across these passes. During the inbound and outbound phases of P4 and P7, the PSP traversed the eastern and western sides of the Sun-Earth line, while in P5 and P8, it was on the far and near sides. At P6, it traveled along the near and eastern sides, and at P9, the western and eastern sides. Meanwhile, STEREO-A was consistently positioned behind Earth, maintaining a quadrature configuration. Lastly, SolO maintained a quadrature with Earth by staying ahead during P5, while in P6 and P7, it was positioned on the far side of the Sun-Earth line. During P8 and P9, it shifted to the east of the Sun-Earth line. SolO was unavailable during P4, as it was not launched until February 11, 2020.

Regarding radial distance, PSP's trajectory varied between approximately $0.4–0.43 \, \mathrm{au}$ at the start of each pass, reaching perihelion distances from $0.13 \, \mathrm{au}$ down to $0.074 \, \mathrm{au}$ in later passes, before extending outward to about $0.48–0.50 \, \mathrm{au}$. STEREO-A maintained a relatively stable heliocentric distance of $0.96–0.97 \, \mathrm{au}$, while SolO’s radial distance varied significantly across passes, ranging from $0.5 \, \mathrm{au}$ to $0.99 \, \mathrm{au}$. Finally, all probes remained close to the ecliptic plane, staying within $\pm 7^\circ$ latitude (not shown here) relative to the solar equator.

\subsection{Calculation of IMF Polarity} \label{subsec:IMFPolarity}

\begin{figure*}[ht!]
\centering
    \includegraphics[width=0.8\textwidth]{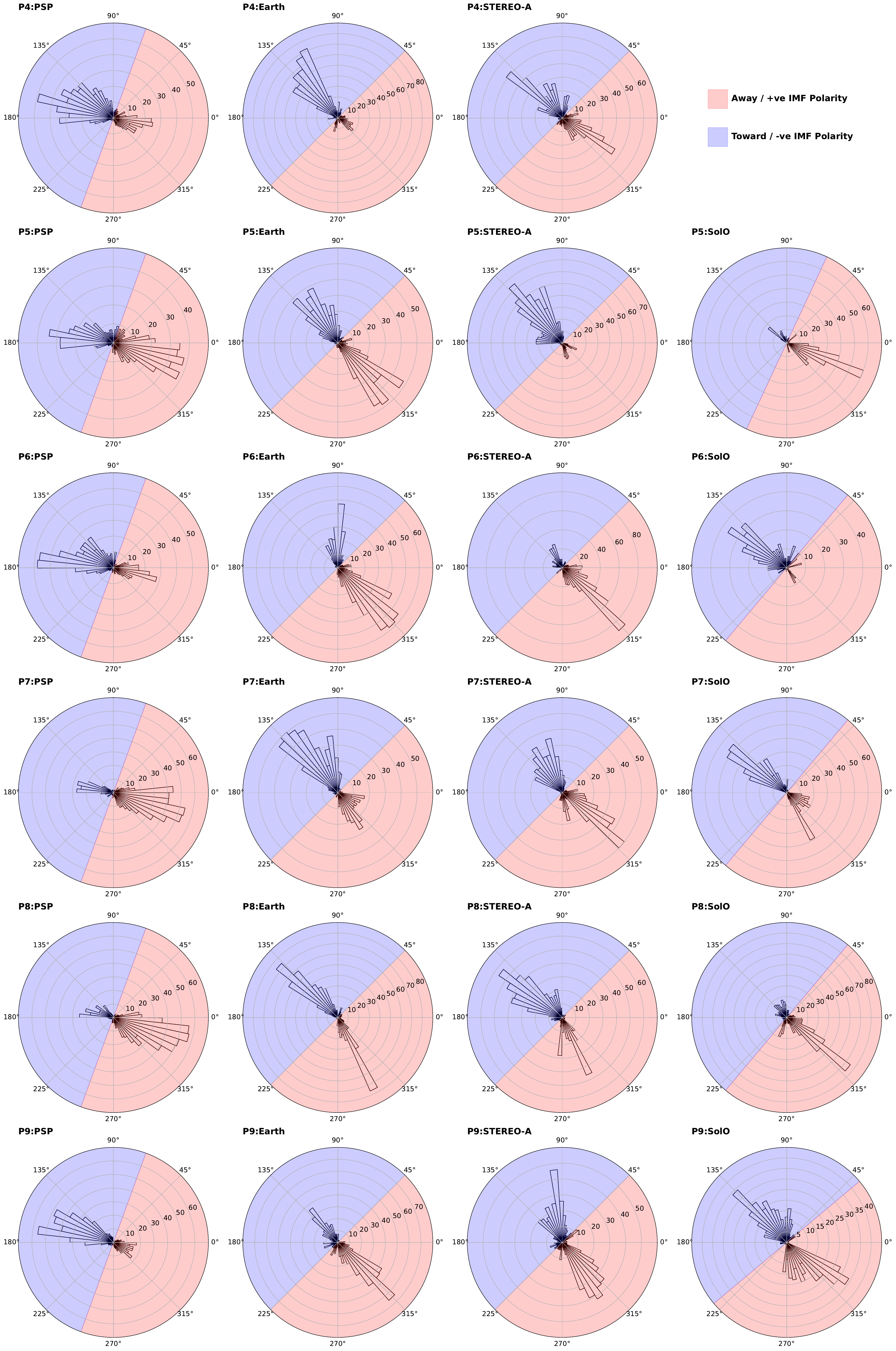}

\caption{Histograms of the observed azimuthal field angle in the RT plane of the RTN coordinate system are presented for different probes (in columns) during PSP perihelion passes four through nine (in rows). The polar axis represents the azimuthal angle($\Phi$), while the radial axis indicates the number of data points, with a bin size of 6 degrees. Where $\Phi = 0$ represents the radial direction away from the Sun (Sun-spacecraft line). Two distinct populations of field angles are evident, with their average field angles differing by approximately 180 degrees. The red and blue shaded regions correspond to positive and negative IMF polarity, indicating the direction away from and toward the Sun, respectively.}

\label{fig:IMFPolarity}
\end{figure*}

The polarity of IMF sectors in our model is determined based on the HCS location in the IHS. Consequently, any magnetic field polarity reversals reconstructed at a given spacecraft position are associated with the HCS crossing observed at that spacecraft. As previously mentioned, we solve a separate level-set advection equation, alongside the MHD equations, to passively propagate the HCS specified at the inner boundary. In this framework, positive (negative) level-set values correspond to positive (negative) IMF polarity. Furthermore, we restore the sign of the magnetic field components $B_R$ and $B_T$, by multiplying them by the IMF polarity derived from the level-set method.

To determine the IMF polarity from in situ SW observations, we utilize the field/spiral angle (\(\Phi\)), which represents the angle between the radial component of the IMF and the radially outward direction (\(\Phi = 0^\circ\)) in the RTN coordinate system. Figure \ref{fig:IMFPolarity} presents the histograms of \(\Phi\) for PSP, Earth, SolO, and STEREO-A probes during P4–P9. The polar axis represents \(\Phi\), and the radial axis indicates the number of data points, with a bin size of \(6^\circ\). Two distinct populations of field angles are evident, with their average field angles differing by approximately \(180^\circ\), which indicates the presence of two separate field directions (towards and away from the sign).

At Earth and STEREO-A, the distribution of the two distinct populations of \(\Phi\) peaked approximately at \(135^\circ\) and/or \(315^\circ\), depending on the dominant angle during that period. Following \cite{MacNeice:2009} (which considered only the near-Earth observations), we assign \(\Phi\) between \(225^\circ\) and \(45^\circ\) to positive IMF polarity, representing the direction away from the Sun. Conversely, the remaining angles are assigned to negative IMF polarity, corresponding to the direction towards the Sun. In Figure \ref{fig:IMFPolarity}, the IMF polarity sectors are shaded blue, indicating the towards polarity (\(-1\)), and red representing the away polarity (\(+1\)). They correspond to the region covering \(\pm 90^\circ\) around the peak field angle, \(\Phi_{\text{p}}\). 

On the other hand, the values of \(\Phi\) at PSP are distributed with peaks approximately at \(160^\circ\) or \(340^\circ\) across all perihelion passes. Accordingly, we assign \(\Phi\) between \(250^\circ\) and \(70^\circ\) to positive IMF polarity, while the remaining angles to negative IMF polarity.

For SolO, the distribution of \(\Phi\) varies slightly among the perihelion passes due to significant changes in SolO’s heliocentric distance during different perihelion passes of PSP. Specifically, at P5, the values of \(\Phi\) are peaked around \(335^\circ\). For P6, P7, and P8, they peaked at approximately \(140^\circ\) or \(320^\circ\), depending on which component is dominant. Lastly, for P9, values of \(\Phi\) are peaked around \(130^\circ\). Hence, we assign field angles within the range of \(\Phi_{\text{p}} \pm 90^\circ\) to the corresponding polarity, as illustrated by the shaded regions in Figure \ref{fig:IMFPolarity}.

\section{Results}\label{sec:Results}

\subsection{2D Distributions of MHD Simulations}

\begin{figure*}[ht!]
\centering
    \includegraphics[width=0.81\textwidth]{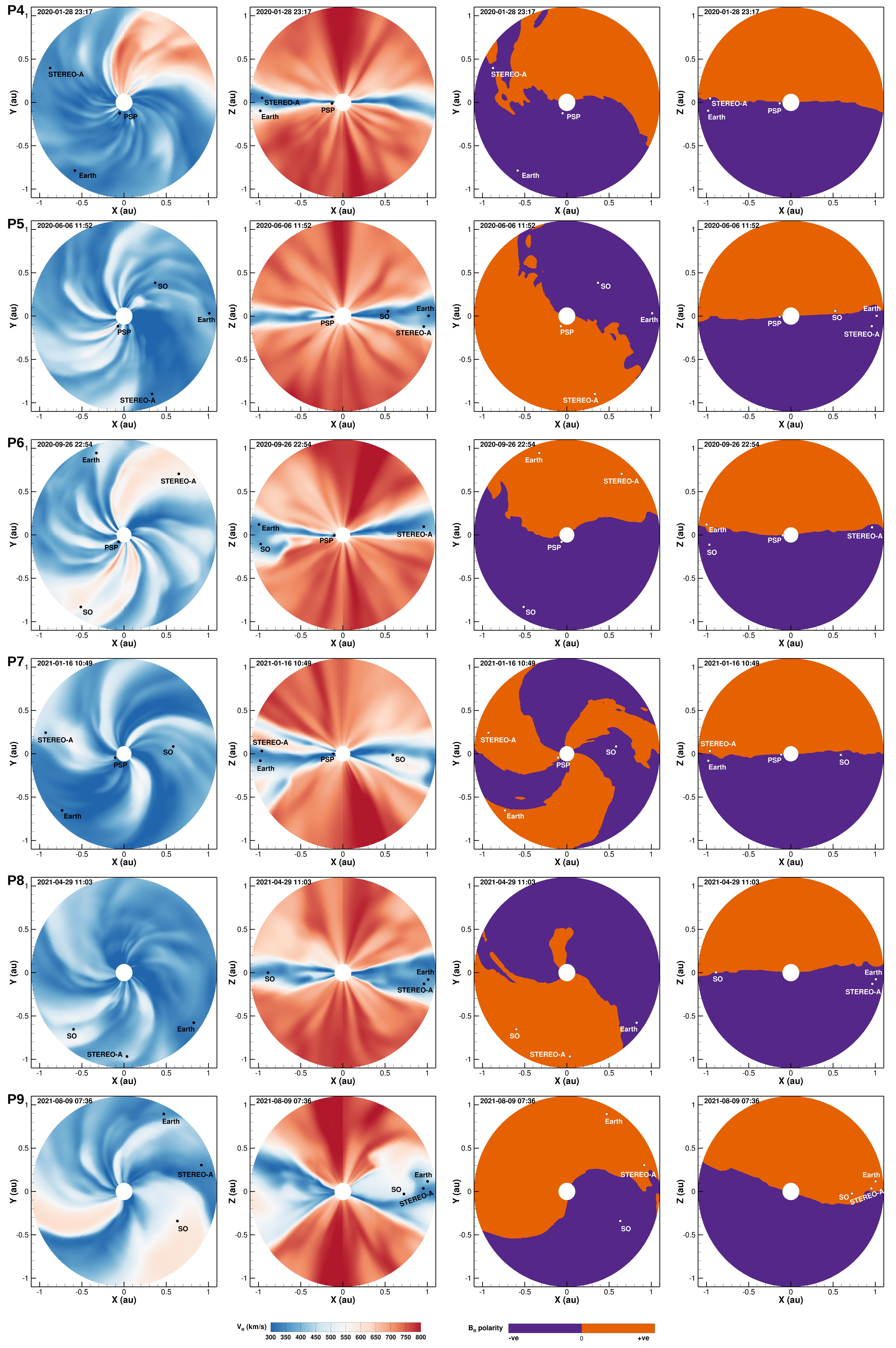}

\caption{Two-dimensional 
cross sections from the 3D MHD simulations for the fourth to ninth PSP perihelia, each row corresponding to a specific perihelion's zeroth realization (R00). The solutions are displayed in the heliospheric coordinate system as described in Subsection \ref{subsec:IHmodel}. The inner boundary is shown with the heliocentric white circle, while the outer boundary is at $1.1 \, \mathrm{au}$. The projected probe positions are labeled. Columns 1 and 2 show $V_R$ in the equatorial ($X$-$Y$ plane) 
and vertical ($X$-$Z$ plane) planes 
, respectively; columns 3 and 4 show the signs 
of $B_R$ in the equatorial and vertical planes, respectively.}

\label{fig:MHD_2D_slices}
\end{figure*}

Figure~\ref{fig:MHD_2D_slices} presents two-dimensional (2D) cross sections from our 3D simulation. In particular, the modeled SW radial velocity component, $V_R$, and the sign of the radial component of the magnetic field, $B_R$, are shown in the equatorial (columns 1 and 3) and vertical (columns 2 and 4) planes. These results correspond to the time intervals around the fourth to ninth PSP perihelia (P4–P9) and are obtained using the zeroth realization (R00). The data are presented in the heliospheric coordinate system, as described in subsection~\ref{subsec:IHmodel}. The inner boundary is shown with the heliocentric white circle, while the outer boundary is located at $1.1 \, \mathrm{au}$. The latitudinal and longitudinal projections of PSP, Earth, STEREO-A, and SolO locations are labeled. PSP is not shown for the P8 and P9 perihelia, as it was located below $21.5\, \mathrm{R_{\odot}}$. The SolO is not displayed for P4 as it had not been launched.

The equatorial slices of $V_R$ exhibit a characteristic Archimedean spiral structure formed by the radial outward motion of the SW plasma combined with solar rotation. These slices also reveal the presence of HSSs, varying in size (longitudinal extent) and speed, along with the source regions (in terms of longitude) of the SW plasma observed by each spacecraft. For example, broader and higher-amplitude HSSs are evident during P4, P6, and P9, while narrower and lower-amplitude faster streams are observed during P5, P7, and P8. These differences reflect the characteristics of their respective source regions. Meanwhile, the equatorial slices of the sign of $B_R$ show typical multiple IMF polarity sectors, with the line (zero contours) separating the colors corresponding to the HCS. This structure reflects the transport of embedded magnetic field lines by the SW plasma in an Archimedean spiral pattern. 

The $V_R$ and $B_R$ solutions in the vertical slices for different perihelia illustrate a gradual increase in the HCS tilt and an expansion of the slow SW region from P4 to P9 as the solar cycle progresses. These slices also highlight that all probes were positioned near the solar equator. However, even slight differences in spacecraft latitudes lead to distinct observations of SW plasma and IMF polarities at each location.

These behaviors of the IHS solution align with the trends observed in the inner boundary conditions shown in Figure ~\ref{fig:bc}. It is important to note that since the SW is already supersonic at the inner boundary, the IHS solutions obtained from the MHD simulation are entirely determined by these boundary conditions. Therefore, we indirectly assess the quality of the boundary conditions by comparing the SW solutions with in situ observations from multiple spacecraft in the following sections.

\subsection{Multi-point Evaluation of the Ensemble Model}

In Figures~\ref{fig:P4} through \ref{fig:P9}, we present a direct comparison of ensemble solutions with in situ SW observations along the trajectories of PSP, Earth, STEREO-A, and SolO during P4 to P9. For each spacecraft trajectory, the panels show, from top to bottom: the hourly averaged magnetic field strength ($B$), the radial components of the magnetic field ($B_R$) and velocity ($V_R$) vectors (with speed ($V$) shown only for STEREO-A), proton number density ($N_p$) and temperature ($T_p$), and IMF polarity. Both the velocity and magnetic field vector components are provided in the RTN coordinate system. The 12 ensemble members of the simulation, the ensemble mean, and the observations are represented in orange, red, and black colors, respectively. The scatter in the orange lines represents the level of uncertainty in the SW simulation due to the uncertainties in the ADAPT model. The grey dotted vertical lines in the panels of $V_R$, $N_p$, and $T_p$ for PSP indicate the intervals when PSP/SPC data are replaced with PSP/SPAN-I observations, while the grey dotted horizontal lines in the $B_R$ and $B$ panels mark the zero line. The grey-shaded intervals represent periods of in situ CME observations from the ICME catalog. In the panels of IMF polarity, the values $\pm 1$ correspond to positive and negative IMF polarities, while the value zero indicates a region of numerical uncertainty (level-set values between $-0.5$ and $0.5$) near the HCS. The ensemble mean of the IMF polarity was calculated by averaging the ensemble level-set values and assigning polarities based on the same criteria. To clearly distinguish the observed IMF polarity from the model ensemble, the observed polarity is plotted as black dots at $\pm 1.1$.

The IMF polarity panels show that spacecraft observed multiple polarity reversals on different time scales. These polarity reversals can be categorized into transient and stable polarity reversals. Specifically, transient polarity reversals are characterized by one or more flips within a few hours (less than a day). In contrast, stable polarity reversals exhibit consistent polarity for at least a full day following the reversal. Stable polarity reversals are attributed to the spacecraft crossing the large-scale HCS structure, which separates opposite heliospheric magnetic sectors and is also referred to as a sector boundary (SB) crossing \citep{Smith:2001}. In contrast, transient polarity reversals may occur due to either multiple crossings of a single rippled or fine-structured HCS or single crossings of multiple surfaces/current sheets by the spacecraft \citep{Neubauer:2008}. Given the resolution of the MHD simulations, it is important to note that the MHD model of the IHS is not expected to reproduce all transient IMF polarity flips. Instead, our model evaluation focuses on its ability to replicate stable polarity reversals/ SB crossings.

The distributions of plasma parameters, on the other hand, exhibit clear signatures of HSSs, characterized by a gradual increase and subsequent decrease in speed and temperature, along with a sharp rise and fall in density at the stream interface. Here, we define an HSS as an enhancement in SW speed of at least $150 \, \mathrm{km\,s}^{-1}$ above a minimum of $300 \, \mathrm{km\,s}^{-1}$, persisting for several days, with peak speeds exceeding $450$–$500 \, \mathrm{km\,s}^{-1}$ and occasionally reaching above $750 \, \mathrm{km\,s}^{-1}$. Since we simulate the SW flow using the MHD model with time-dependent boundary conditions, HSSs and associated SIRs are developing self-consistently. Below we evaluate the performance of our ensemble model in reproducing the arrival time and overall structure of HSSs observed at different spacecraft.

In the following subsections, we first describe the large-scale structures observed at each spacecraft in the IHS and compare them with our ensemble simulation results for each PSP perihelion pass. Specifically, we assess the average performance of the simulation in capturing the observed structures within the range of uncertainties. This is done by visually examining the time-series plots to determine whether the ensemble members encompass the observations. Accurate reconstruction of SW structures from the simulation results enhances our understanding of in situ observations, while analyses of discrepancies help identify limitations in the current simulation setup and provide guidance for improving future modeling approaches.

\subsubsection{Perihelion Pass 4 (CRs 2226-2227)}

\begin{figure*}[ht!]
\centering
\includegraphics[width=\textwidth]{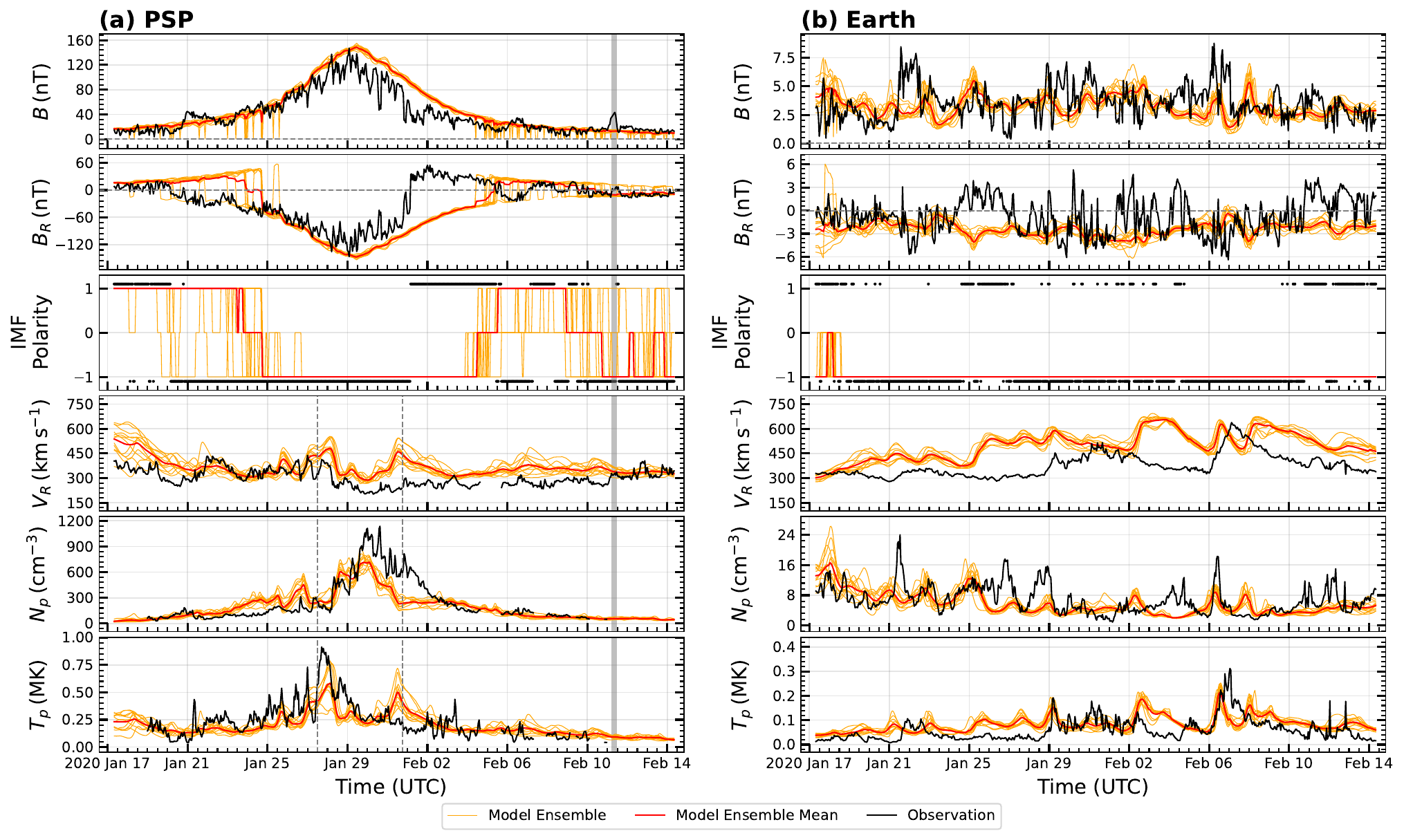}

\caption{Comparison of the numerical solutions with in situ observations along the trajectories of PSP (a), and Earth (b) during P4, covering the period from January 17 to February 14, 2020. For each panel, from top to bottom, the magnetic field strength ($B$) and its radial component ($B_R$) in $\, \mathrm(nT)$, IMF polarity, radial velocity component ($V_R$) in $\, \mathrm{km}\,\mathrm{s}^{-1}$, plasma number density ($N_p$) in $\, \mathrm{cm}^{-3}$, and temperature ($T_p$) in $\, \mathrm(MK)$. The 12 ensemble members of the simulation, the ensemble mean, and the observations are represented in orange, red, and black colors, respectively. The grey dotted vertical lines in the panels of $V_R$, $N_p$, and $T_p$ for PSP indicate the intervals when PSP/SPC data are replaced with PSP/SPAN-I observations, while the grey dotted horizontal lines in the $B_R$ and $B$ panels mark the zero line. The grey-shaded intervals represent periods of in situ CME observations from the ICME catalog.}
\label{fig:P4}
\end{figure*}

\begin{figure}[ht!]
\centering

\includegraphics[width=0.49\textwidth]{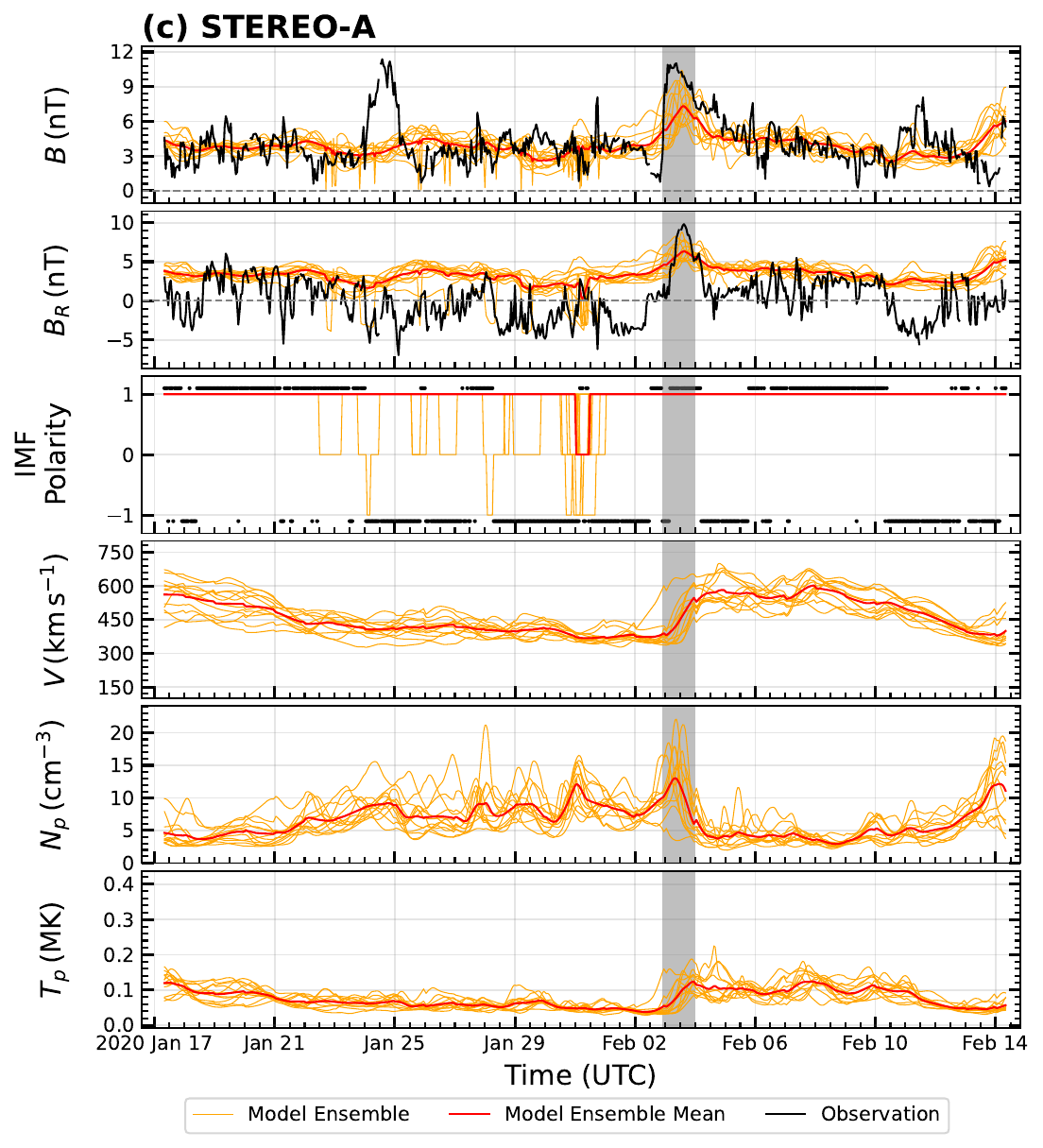}

\caption{Comparison of the numerical solutions with in situ observations along the trajectory of STEREO-A during P4, covering the period from January 17 to February 14, 2020.  Same format as Figure \ref{fig:P4}. The panel of $V_R$ is replaced by SW speed, $V$, due to the unavailability of angular components of the velocity. Note: plasma data for STEREO-A were unavailable during this period. }
\label{fig:P4_1}
\end{figure}

As shown in Figure~\ref{fig:P4}(a), PSP observed a gradually increasing (decreasing) $B$ during the inbound (outbound) phase of P4, reaching a heliocentric distance of approximately $0.1\, \mathrm{au}$ and recording the amplitude of $\sim \! \! \! 140 \, \mathrm{nT}$ on January 29, 2020, during its fourth perihelion. The $B_R$ and IMF polarity panels indicate that PSP predominantly traversed the negative polarity sector, experiencing multiple polarity reversals during P4, when  PSP observed SB crossings on January 20, February 1, 5, 7, and 8, 2020.

On the other hand, the ensemble mean of the simulation successfully reconstructed the overall trend and amplitude of $B$, but overestimated it by $\sim \! \! \! 40 \, \mathrm{nT}$ two days before and six days after perihelion. This overestimation primarily arises from the overestimation of $B_R$. Additionally, discrepancies of $\sim \! \! \! 60\, \mathrm{nT}$ and $\sim \! \! \! 120\, \mathrm{nT}$ exist in $B_R$ during January 20–25 and February 1–4, 2020, respectively. These discrepancies are related to the fact that the simulation missed the SB crossings on January 20 and February 1, which resulted in the incorrect reconstruction of the IMF polarity during those periods. While the ensemble mean of the simulation missed most of the SB crossings, at least one ensemble member captured all SB crossings except for the one on February 1, 2020. Note that the largest ensemble spread in $B_R$ originates from the polarity.

Furthermore, PSP consistently observed slow SW in the range of $200 \,\mathrm{km}\,\mathrm{s}^{-1}$ to $450\,\mathrm{km}\,\mathrm{s}^{-1}$ throughout the perihelion pass, with only slight variations of $\sim \! \! \! 150\ \mathrm{km}\,\mathrm{s}^{-1}$ in the flow speed. As the spacecraft approached perihelion, a gradual increase in proton density was observed, reaching approximately $1100\, \mathrm{cm}^{-3}$, followed by a gradual decrease, reflecting its dependence on heliocentric distance. The proton temperature measured by PSP increased steadily, reaching a peak of about $0.9\, \mathrm{MK}$ on the day before perihelion, and remained below $0.3\, \mathrm{MK}$ during the outbound phase.

In comparison, the ensemble mean of our simulation reproduced the general trend in $V_R$, but overestimated it by $\sim \! \! \! 100\ \mathrm{km}\,\mathrm{s}^{-1}$ throughout the duration, except during January~21--27, 2020, and February~11--14, 2020. Regarding proton density and temperature, the ensemble simulation reproduced the observed values well within the range of uncertainty for most of the period. However, during the four days following perihelion, discrepancies of $\sim \! \! \! 300\, \mathrm{cm}^{-3}$ in proton density and $\sim \! \! \! 0.1\, \mathrm{MK}$ in temperature were noted. These deviations are attributed to the simulation incorrectly producing a fast, hot, and low-density wind originating from the southern hemisphere, which was not observed. Regarding the ensemble spread, significant variability was present throughout the period, with maximum spreads of approximately $300 \,\mathrm{km}\,\mathrm{s}^{-1}$  in speed, $300\, \mathrm{cm}^{-3}$ in density, and $0.5\, \mathrm{MK}$ in temperature.

Figure~\ref{fig:P4}(b) shows that $B$ and $|B_R|$ at Earth were significantly smaller than those at PSP, which is consistent with the expected decrease in magnetic field strength with increasing heliocentric distance. However, similar to PSP, Earth experienced several transient IMF polarity reversals throughout period P4, along with a few stable SB crossings on January 18, 24, and 27, 2020, as well as February 10, 2020. In contrast, the simulation indicates that Earth remained in a negative polarity region throughout the interval and did not capture the SB crossings observed between January 24-27 and February 10-14, 2020. While the ensemble simulation exhibited some spread of $\sim \! 3 \, \mathrm{nT}$ in $B$ and $|B_R|$, no spread was observed in IMF polarity, except during the HCS crossing on January 18, 2020.

Earth predominantly observed slow SW with $V_R$ of $\sim \! \! 300\, \mathrm{km}\,\mathrm{s}^{-1}$ during the first half of P4, followed by the two HSSs centered on January 31, 2020, and February 7, 2020, in the second half. The negative/inward IMF polarity of the HSSs indicates that these streams originated from coronal holes in the southern hemisphere. The first HSS exhibited a complex structure with multiple peaks reaching an amplitude of $\sim \! \! \! 500 \, \mathrm{km}\,\mathrm{s}^{-1}$, while the second displayed a typical single, sharp peak with an amplitude of $\sim \! \! \! 650 \, \mathrm{km}\,\mathrm{s}^{-1}$ and a relatively narrower width compared to the first. 

In contrast, discrepancies of approximately $200\, \mathrm{km}\,\mathrm{s}^{-1}$, $250\, \mathrm{km}\,\mathrm{s}^{-1}$, and $150 \, \mathrm{km}\,\mathrm{s}^{-1}$ were observed between the ensemble mean of the simulation and the observations during January 25-29, 2020, February 2–6, 2020, and February 8–11, 2020, respectively. These discrepancies are attributed to inaccuracies in simulating the HSS and the associated SIRs. Specifically, the ensemble simulation did not capture the single peak structure and $V_R$ during the trailing edge of the second HSS. For the multi-peaked HSS, the simulation could not accurately represent the rising and trailing edges, except for a portion of the peak. The average SW speed was significantly overestimated across all ensemble members. Furthermore, the simulation produced faster streams during January 25–29, 2020, and February 2–6, 2020, which were absent in the observations.

The observations showed multiple density enhancements, some of which, centered on January 28 and February 6, 2020, coincided with the interfaces of the SIRs formed by the first and second HSSs—i.e., the compression regions. Others were associated with HCS crossings or stand-alone high-density structures. In contrast, while the ensemble simulation reasonably reproduced the overall density trend, it did not capture the density enhancements in the slow SW. Larger discrepancies arose due to inaccuracies in reconstructing the observed HSSs and introducing unobserved ones. Regarding proton temperature, the simulation reproduced the general trend in the first HSS reasonably well, within the range of uncertainty. For the second HSS, the ensemble simulation accurately captured the stream interface during the rising and trailing edges, but not the amplitude. The typical levels of uncertainty at Earth for radial velocity, density, and temperature are $100\, \mathrm{km}\,\mathrm{s}^{-1}$, $8\, \mathrm{cm}^{-3}$, and $0.05 \, \mathrm{MK}$, respectively. No significant differences were observed in the ensemble spread across the period or variables.

In Figure~\ref{fig:P4}(c), we present our simulation results alongside STEREO-A observations. Plasma data for this period was unavailable; therefore, only the simulation results for the plasma variables are shown. We can see that $B$ and $|B_R|$ are of the same order as observations at Earth, consistent with STEREO-A’s heliocentric distance near $1 \, \mathrm{au}$. The ensemble simulation captured the overall trend in $B$, except during January 24–25, 2020, where it missed an enhancement with an amplitude of approximately $6\, \mathrm{nT}$. $B_R$ and IMF polarity observations indicate that STEREO-A experienced frequent transient polarity reversals and stable SB crossings on January 23, February 6, and February 10, 2020. In contrast, the simulation suggests that STEREO-A remained predominantly in the positive polarity region throughout this period, without reproducing any HCS crossings. Some ensemble members did capture a few polarity reversals between January 22 and January 31, though these were limited to transient flips.

Although plasma data is unavailable to evaluate the simulation comprehensively, the level of uncertainty at STEREO-A can be inferred from the ensemble spread, which shows a relatively larger spread compared to Earth and PSP. 

\subsubsection{Perihelion Pass 5 (CRs 2231 -2232)}

\begin{figure*}[ht!]
\centering
\includegraphics[width=\textwidth]{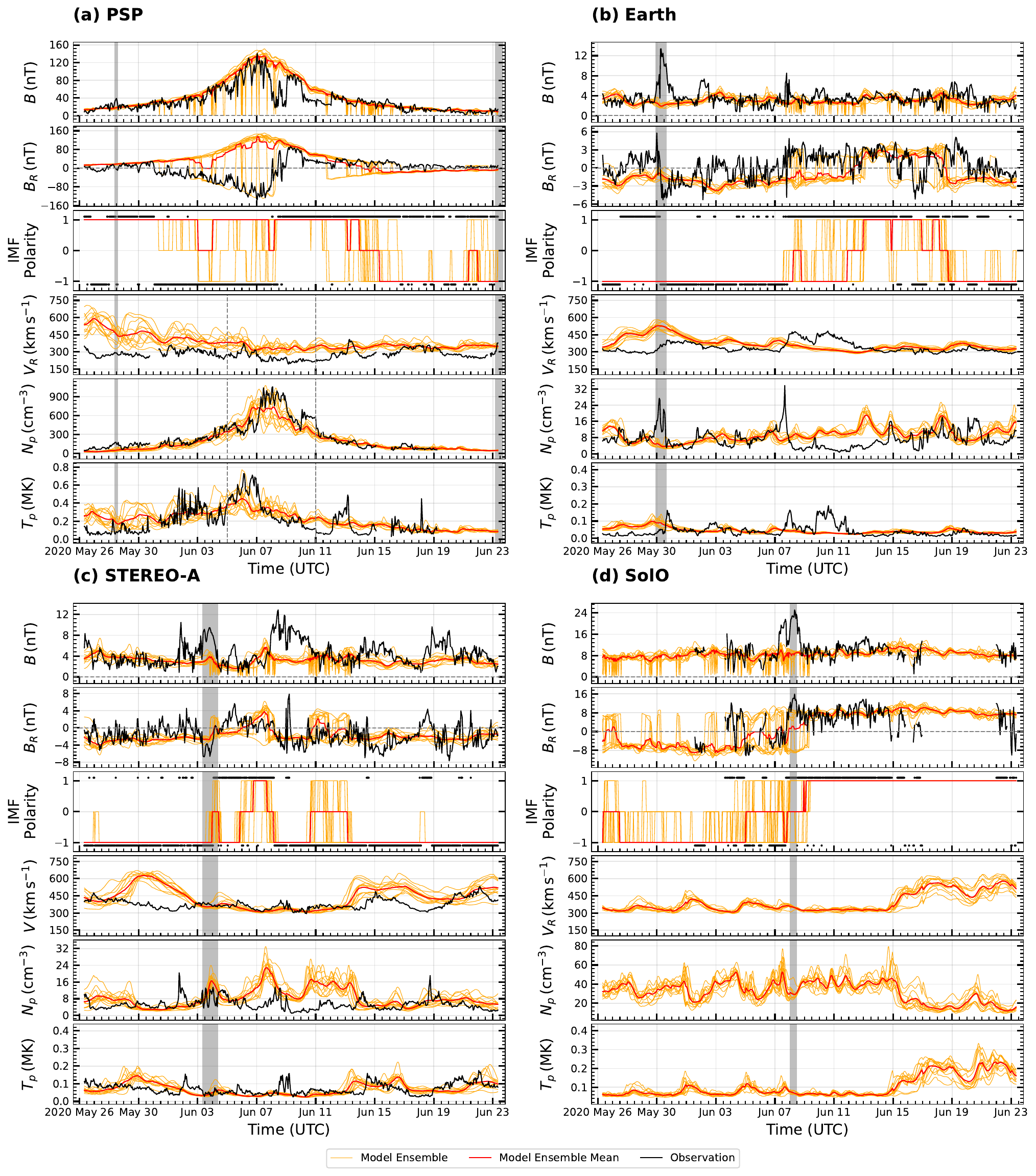}

\caption{Comparison of the ensemble simulation results with in situ observations along the trajectories of PSP (a), Earth (b), STEREO-A (c), and SolO (d) during P5, from May 26 to June 23, 2020. Same format as Figure \ref{fig:P4}. Note: plasma data for SolO were unavailable during this period.}

\label{fig:P5}
\end{figure*}

As shown in Figure~\ref{fig:P5}(a), during P5, PSP observed the distributions of $B$ and $|B_R|$ similar to those of P4, with amplitudes reaching approximately $\sim \! \! 140\,\mathrm{nT}$ on June 7, 2020, when it also reached a heliocentric distance of $0.1\, \mathrm{au}$ during its fifth perihelion. However, the $B_R$ and IMF polarity panels indicate that PSP predominantly traveled through the positive polarity sector during P5, in contrast to the negative sector observed during P4. Several polarity reversals were identified, particularly on May 26, 27, 31, and June 8, 2020, coinciding with the SB crossings. On June 8, 2020, the day following perihelion, PSP crossed the HCS twice, leading to the observed double dip in $B$. In addition, the frequent polarity reversals during the outbound phase reflect that the spacecraft traversed in the SW very close to the HCS. 

On the other hand, the simulation reconstructed the distributions of $B$ and $|B_R|$ reasonably well within the range of uncertainties. However, some ensemble members produced multiple dips in $B$ approaching zero, attributed to polarity reversals in the simulation. Notably, in observations, $B$ does not typically drop to zero during polarity reversals caused by HCS crossings; only minor dips are generally observed. While at least one ensemble member reproduced the SB crossing on June 8, 2020, the simulation captured only one crossing that day rather than two consecutive flips within a short interval. Additionally, the simulation missed the HCS crossing on May 26–27 and suggested several unobserved HCS crossings on June 15–20, 2020.

Furthermore, PSP was predominantly immersed in the slow SW, with radial speeds ranging approximately between $200\, \mathrm{km}\,\mathrm{s}^{-1}$  and $400 \, \mathrm{km}\,\mathrm{s}^{-1}$  exhibiting minimal variability throughout P5. the simulation successfully captured this behavior, although it systematically overestimated the $|V_R|$ by at least $100\, \mathrm{km}\,\mathrm{s}^{-1}$. A more than $150 \, \mathrm{km}\,\mathrm{s}^{-1}$ discrepancy was observed during May 26–28, when the simulation missed the HCS crossings and instead produced an unobserved HSS. Regarding proton density, PSP observed a trend similar to that at P4, with amplitudes of $\sim \! \! \! 1000 \, \mathrm{cm}^{-3}$, and the simulation reconstructed these observations reasonably well within the range of uncertainties. Finally, for proton temperature, the simulation accurately captured the overall trend within the level of uncertainties, except during May 26–28 and June 10–11, 2020, where the simulation slightly overestimated it by $\sim \! \! 0.1 \, \mathrm{MK}$.

Figure~\ref{fig:P5}(b) shows that a mean $B$ of $\sim \! \! 4\,\mathrm{nT}$ was observed at Earth, with slight variations during P5. Significant variations of $B_R$ were observed, primarily due to multiple polarity reversals. Earth spent nearly equal amounts of time in both polarity sectors, largely because the Earth’s latitude was very close to the heliographic equator (ranging from $-1 ^{\circ}$  to $2 ^{\circ}$) and the HCS was nearly parallel to the equator during P5. In contrast, the simulation indicated that Earth remained predominantly in the negative polarity region for the first twelve days and did not capture the polarity reversals on May 27 and 30, 2020. Nevertheless, the simulation accurately reproduced the HCS crossings and $|B_R|$ for the remainder of the period, staying well within the range of uncertainties. Furthermore, the ensemble simulation and observations agreed well for $B$.

Earth was predominantly immersed in the slow SW during P5 and observed an HSS with double peaks between June 7–13, 2020. While the simulation successfully reconstructed the general slow wind profile, it completely missed the observed HSS and suggested an HSS during May 26–31, 2020, which was not present. This mismatch resulted in $\sim \! \! \! 150 \, \mathrm{km}\,\mathrm{s}^{-1}$ discrepancies in both cases. However, the simulation reproduced the proton density reasonably well within the level of uncertainties, except when the missed HSS led to a discrepancy of about $16 \, \mathrm{cm}^{-3}$, in the compression region caused by the interaction of the HSS with the slow wind. Finally, the discrepancies and agreements between the simulation and the observations in proton temperature followed a pattern similar to that of $V_R$.

As shown in Figure~\ref{fig:P5}(c), the distribution of $B$ observed by STEREO-A differed from that at Earth, particularly in terms of variations caused by magnetic field enhancements throughout P5. While the simulation agreed well with the mean $B$, it missed several enhancements. Furthermore, STEREO-A spent most of the period in the negative IMF polarity region, with brief HCS crossings into the positive sector and back during June 3–7 and on June 18, 2020. In comparison, the simulation accurately reproduced most IMF polarities and captured the arrival times of the HCS crossings on June 3 and 7, 2020, within the uncertainty range. However, the simulation produced an additional unobserved HCS crossing between June 10–13, 2020, although at least one ensemble member accurately maintained the negative polarity during this period.

STEREO-A predominantly observed slow SW and two HSSs centered on June 15 and 21, 2020. The simulation successfully captured both HSSs, though their arrival times were slightly early. However, for the HSS centered on June 15, 2020, the speed was overestimated by $\sim \! \! \! 100 \,\mathrm{km}\,\mathrm{s}^{-1}$ in the rarefaction region. Otherwise, the simulation reproduced most of the slow wind data while suggesting an unobserved HSS centered on May 31, 2020, resulting in a discrepancy of about $225\, \mathrm{km}\,\mathrm{s}^{-1}$. Regarding proton density, the simulation accurately reproduced observations for most of the interval, but overestimated values between June 6 and 13, 2020, by approximately $9\, \mathrm{km}\,\mathrm{s}^{-1}$. Additionally, for temperature, the simulation reproduced the baseline and enhancements associated with the compression region created by HSSs during the final week of the perihelion pass, albeit with slight differences in the arrival times and magnitudes. However, the simulation missed the high-temperature, high-density structures that peaked on June 2 and 9, 2020. Notably, for SW speed and temperature, the ensemble spread during the slow wind was relatively smaller than during the HSS, whereas, for density, the opposite trend was observed.

Figure~\ref{fig:P5}(d) shows that SolO observed a mean $B$ of approximately $\sim \! 8\,\mathrm{nT}$, higher than that at Earth and STEREO-A, due to SolO’s closer heliocentric distance during P5. The $B_R$ and IMF polarity data indicate that SolO was in the positive polarity sector during the second half of the period, while multiple polarity reversals during the first half suggest that SolO was traversing near the HCS. In comparison, the simulation successfully reproduced the IMF polarity, $|B_R|$, and $B$, well within the uncertainty range. Although SW plasma data from SolO were unavailable, the simulation results are presented here for reference.

\subsubsection{Perihelion Pass 6 (CRs 2235-2236)}

\begin{figure*}[ht!]
\centering
\includegraphics[width=\textwidth]{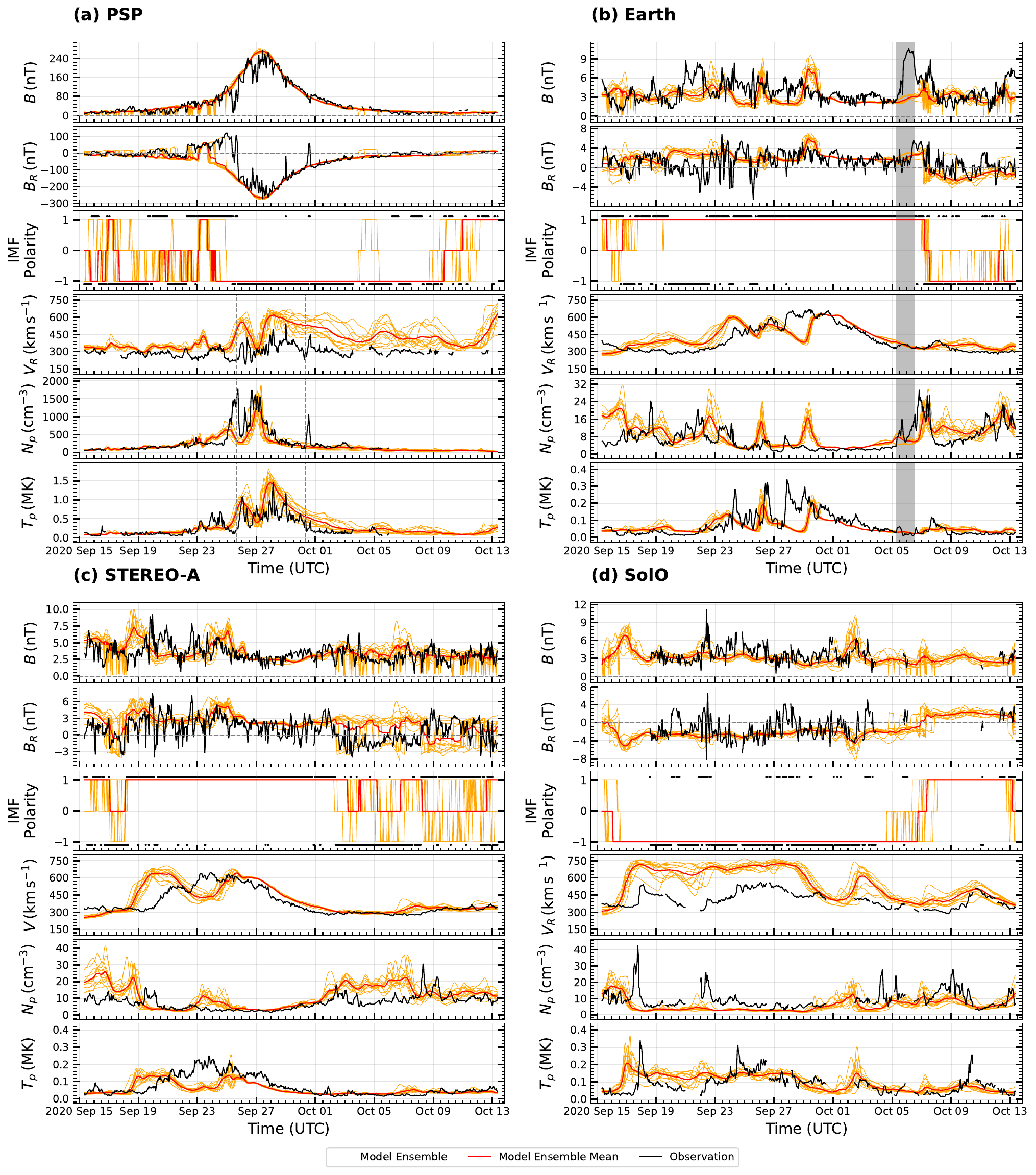}
\caption{Comparison of our ensemble simulation with in situ observations along the trajectories of PSP (a), Earth (b), STEREO-A (c), and SolO (d) during P6, from September 15 to October 13, 2020. The format is as in Figure \ref{fig:P4}.}
\label{fig:P6}
\end{figure*}

As shown in Figure~\ref{fig:P6}(a), during P6, PSP observed $B$ and $|B_R|$ profiles similar to those at previous perihelia. However, with amplitudes reaching approximately $\sim \! \! \! 260\,  \mathrm{nT}$ on September 27, 2020, at a heliocentric distance of $0.095 \, \mathrm{au}$, PSP was closer to the Sun during its sixth perihelion than at the earlier perihelia. PSP traversed the SW very close to the HCS during P6, except near perihelion, as indicated by multiple IMF polarity reversals. However, it remained mostly in the negative sector for the remainder of the period.

In comparison, the ensemble simulation captured the general trend and amplitude of $B$, although it missed some intensity dips around perihelion. The simulation successfully reproduced all HCS crossings during the inbound phase, at least by a single ensemble member. However, the timing errors are apparent in some simulated HCS crossings, such as the one that occurred two days before perihelion. Additionally, the simulation reconstructed the predominantly negative polarity during the outbound phase, but missed a few HCS crossings between October 5–8, 2020, while successfully capturing the rest. Although the overall distribution of $B_R$ was reconstructed well within the range of uncertainties, a discrepancy of $\sim \! \! 200\, \mathrm{nT}$ was observed during September 25–26, 2020, resulting from the HCS crossing one day earlier in the simulation compared to the observations.

Regarding plasma properties, PSP observed slow SW with speed in the range of  $150-450\, \mathrm{km}\,\mathrm{s}^{-1}$ throughout P6, except for a few minor speed enhancements ($> \! \! \! 450 \, \mathrm{km}\,\mathrm{s}^{-1}$) near perihelion. In contrast, the ensemble simulation reproduced the variations with slight timing mismatches during the inbound phase up until two days before perihelion. Moreover, it incorrectly suggested two unobserved HSSs near perihelion and systematically overestimated the speed of the very slow wind ($< \! \!300 \ \mathrm{km}\,\mathrm{s}^{-1}$), similar to previous perihelia. During the outbound phase, all ensemble members consistently exceeded the observed values and suggested multiple unobserved HSSs. Notably, the ensemble simulation exhibited a larger scatter of $\sim \! \! 300\, \mathrm{km}\,\mathrm{s}^{-1}$ during the outbound phase, compared to about $75 \, \mathrm{km}\,\mathrm{s}^{-1}$ for the inbound phase.

Further, PSP observed two high-density structures, with amplitudes exceeding $1500\, \mathrm{cm}^{-3}$ near perihelion on September 25 and 27, 2020. The ensemble simulation successfully captured the latter enhancement within the uncertainty range, but underestimated the amplitude and showed a timing mismatch for the former structure. Additionally, the simulation did not reproduce the sharp spike-like enhancements in density and velocity coinciding with the IMF polarity change on September 30, 2020, as the HCS crossing responsible for this was missed in the simulation. Finally, the ensemble simulation successfully reconstructed proton temperature within the level of uncertainty, except for a slight overestimation of a small enhancement around September 24, 2020.

Figure~\ref{fig:P6}(b) shows that positive IMF polarity was predominantly observed at Earth, with a relatively brief interval of negative polarity during P6. The spacecraft experienced SB crossings on September 16, 19, 22, and October 7, 2020, along with numerous transient polarity reversals. Otherwise, distributions of  $B$ and $B_R$ exhibited typical magnitudes between $0$ and $10\, \mathrm{nT}$, with relatively more pronounced variations. In contrast, the ensemble simulation reproduced the overall trend in the $B$ and $B_R$ reasonably well, except during September 19–22, 24, and 28, 2020, when the specific variations were not captured. Most of the SB crossings were reconstructed by at least one ensemble member. However, all realizations of the simulation missed the SB crossing on September 19, 2020, maintaining the positive polarity sector instead of transitioning to the negative one, which resulted in discrepancies in simulating the $B_R$. Nevertheless, the ensemble members reproduced $B_R$ more accurately than during the previous two perihelion passes.

Regarding SW plasma, observations show that Earth was predominantly immersed into a broad HSS with a peak speed of $650\, \mathrm{km}\,\mathrm{s}^{-1}$ that lasted $\sim \! \!12$ days and was centered around September 29, 2020, featuring a relatively complex structure. Based on the positive IMF polarity, the HSS likely originated from a source region in the northern hemisphere of the solar corona. On the other hand, the simulation captured the overall trend of the $V_R$, but did not reproduce the complex peak of the HSS. Instead of reconstructing a single broad HSS, the simulation produced three fragmented fast streams intermingled with slow SW, resulting in a speed discrepancy of about $200\, \mathrm{km}\,\mathrm{s}^{-1}$ at the center of the HSS. However, the simulation successfully reconstructed the trailing edge of the HSS, while indicating an earlier rise and faster wind near the beginning of the stream interface. Notably, the ensemble spread along the time axis suggested an arrival time error of approximately $1.5$ days for the simulated HSS. 

While multiple density enhancements were observed, particularly in the compression regions at the stream interfaces, the simulation did not accurately capture the density variations and amplitudes in these regions or during the leading edge of the HSS. However, densities in the rarefaction regions were reproduced relatively well. Finally, the simulation successfully replicated most of the observed variations in proton temperature, except for the peaks associated with the leading edge and the center of the HSS. The largest discrepancies,$\sim \! \! 12\, cm^{-3}$ for density and $0.3\, \mathrm{MK}$ for temperature occurred at the center of the HSS.

Figure~\ref{fig:P6}(c) shows that STEREO-A observed a plasma and magnetic field environment resembling that of Earth during P6, but occurring four days earlier in time. This similarity is evident in the SW speed and the corresponding magnetic field distributions, as the same broad HSS was observed at both locations due to corotation. However, the proton temperature and density distributions differed between the two locations. The ensemble simulation successfully reconstructed the $B$ and $B_R$ distributions while accurately capturing all IMF polarity reversals associated with SB crossings, such as those on September 16, 18, and October 2, 8, and 12, 2020, within the range of uncertainties. Although the ensemble mean of the simulation showed opposite polarities on October 2-12, 2020, at least a few ensemble members successfully reconstructed the correct HCS crossings.

The simulation reproduced the overall SW speed profile for most of the period, except for September 19–25, 2020. During this time, a discrepancy of $\sim \! \! 300\, \mathrm{km}\,\mathrm{s}^{-1}$ during the leading edge of the HSS arose from the early arrival of the HSS in the simulation, which was 1.5 days ahead of the observations. In contrast, the discrepancy of about $150 \, \mathrm{km}\,\mathrm{s}^{-1}$ around the center of the HSS due to the simulation of double fast streams on September 20 and 26, 2020, whereas observations indicated a single broad stream centered on September 24, 2020. Additionally, the simulation slightly overestimated the speed ($50 \, \mathrm{km}\,\mathrm{s}^{-1}$) in the trailing edge of the HSS, consistent with its performance at Earth.

Regarding proton density, the ensemble simulation suggested two distinct density enhancements due to compression regions centered on September 18 and 23, 2020, instead of a single enhancement observed on September 20. Furthermore, the simulation systematically overestimated the density by $\sim \! \! 10 \mathrm{cm^{-3}}$ between October 3–7, 2020, and during the first two days of P6. Finally, the reconstruction of the proton temperature closely resembled the behavior for the simulated SW speed, except during September 20–26, 2020, where the simulation underestimated the temperature in contrast to its overestimation of the speed.

Figure~\ref{fig:P6}(d) shows that SolO mostly observed negative IMF polarity, with multiple short-term polarity reversals throughout the perihelion pass, indicating the spacecraft's proximity to the HCS. Unfortunately, magnetic field data were unavailable during the initial three and final nine days of P6. Nevertheless, the range of $B$ and $|B_R|$ observed by SolO was similar to those at Earth and STEREO-A, as all three spacecraft were orbiting near 1 $au$ during P6. In comparison, the ensemble simulation reconstructed the overall trend in $B$ but did not capture the variations in $B_R$. Notably, the simulation successfully reproduced the predominant negative IMF polarity. However, it did not capture the transient polarity reversals, suggesting that the simulated HCS was not as close to PSP as implied by observations.

Plasma observations at SolO indicate that three HSSs of varying complexity were observed during P6. The first two HSSs, centered on September 18, 2020, and September 26, 2020, originated from the southern hemisphere of the solar corona, with the former exhibiting a sharp rise and the latter displaying a more gradual rise. Meanwhile, the third HSS, observed on October 11, 2020, originated from the northern hemisphere of the solar corona and featured a step-like rise, with all three events concluding in a smooth decline. On the other hand, the ensemble simulation reconstructed multiple HSSs, but was unable to accurately capture the arrival times, amplitudes, and structural complexities of the first two HSSs. Specifically, a discrepancy of $\sim \! \! 300 \, \mathrm{km}\,\mathrm{s}^{-1}$ arose due to the early arrival of the first HSS by about one day. Additionally, an error of about $200 \, \mathrm{km}\,\mathrm{s}^{-1}$ was observed during September 19–29, caused by the overestimation of amplitudes. Although the model produced an HSS on October 3, 2020, that was not observed by SolO, it closely reproduced the third HSS on October 11, 2020, within the uncertainty range.

Furthermore, the simulation consistently underestimated proton density throughout the period and did not capture the density enhancements observed in the compression regions around the stream interfaces of each HSS and at the HCS crossings. Instead, it produced an extraneous density peak corresponding to the incorrectly reconstructed HSS on October 3, 2020. Finally, the ensemble simulation successfully reproduced the amplitude of proton temperature for the first HSS, with an early arrival by one day. For the remaining HSSs, the simulation reproduced the general temperature trend, but did not capture their amplitudes. Interestingly, the errors in reconstructing temperature variations were relatively smaller compared to those for $V_R$.

\subsubsection{Perihelion Pass 7 (CRs 2239-2240)} 

\begin{figure*}[ht!]
\centering
\includegraphics[width=\textwidth]{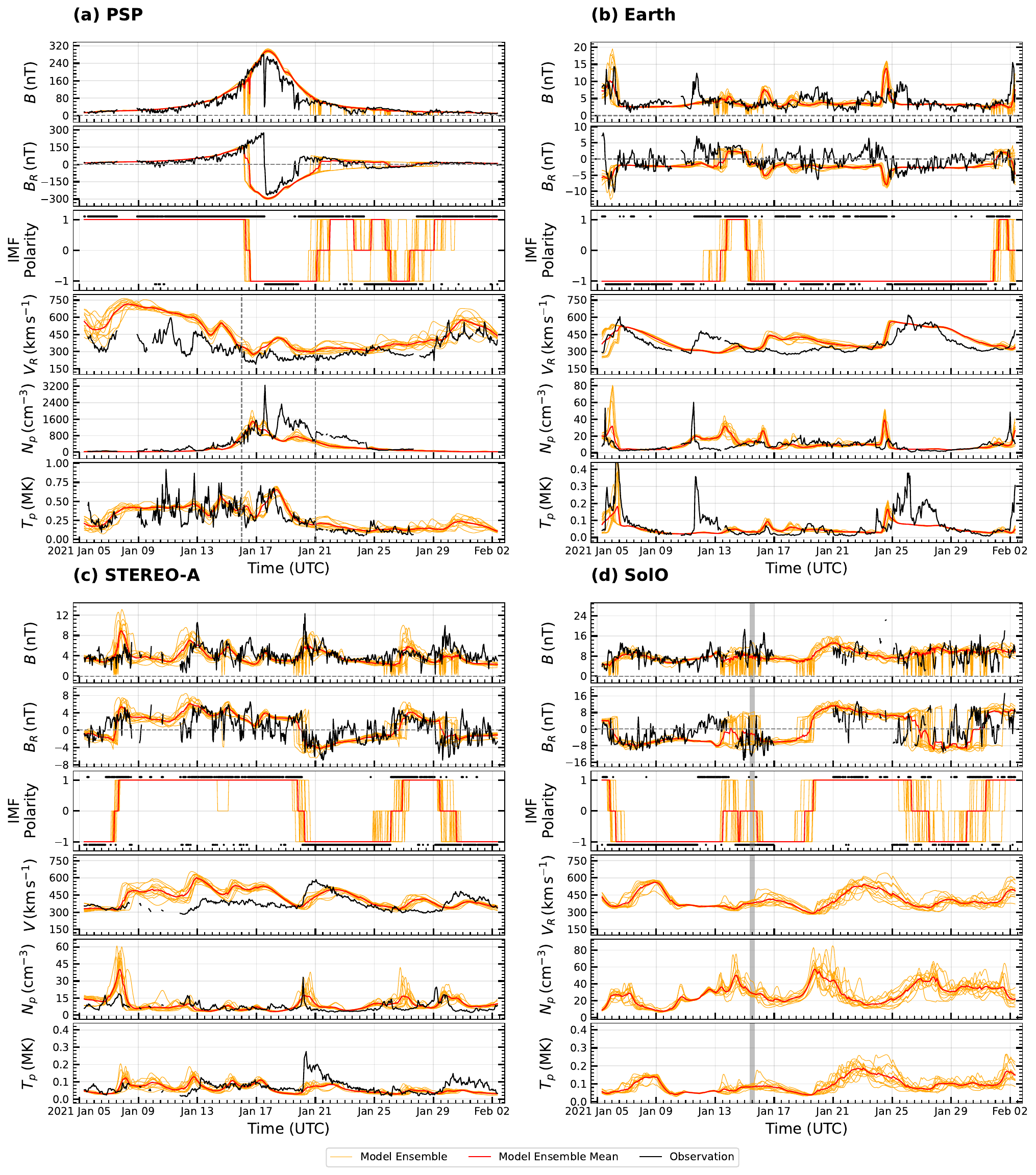}

\caption{Comparison of ensemble simulations with in situ observations along the trajectories of PSP (a), Earth (b), STEREO-A (c), and SolO (d) during P7, from January 05 to February 02, 2021. The same format is the same as in Figure \ref{fig:P4}. Note: Plasma data for SolO were unavailable during this period.}

\label{fig:P7}
\end{figure*}

As shown in Figure~\ref{fig:P7}(a), PSP remained in the positive polarity sector during the inbound phase of P7 and experienced an SB crossing from positive to negative polarity precisely on the perihelion day, January 17, 2020. Subsequently, the spacecraft observed a three-sector structure, with additional SB crossings occurring on January 20, 24, and 27, 2020. The amplitudes of $B$ and $B_R$s were comparable to those observed during the previous perihelion pass, as PSP followed a similar trajectory in P7 as in P6. In contrast, while the simulation reproduced the overall trends in $B$ and $|B_R|$ reasonably well, it systematically overestimated them four days after the perihelion in the outbound phase. Nevertheless, the ensemble simulation successfully reconstructed all SB crossings observed by PSP, albeit with mismatches in the crossing times. For instance, a discrepancy exceeding $300 \, \mathrm{nT}$ during January 16–17, 2020, arose because the HCS crossing occurred one day earlier in the simulation. Similarly, a discrepancy of $\sim \! \! 150 \, \mathrm{nT}$ around January 20, 2020, was due to a one-day delay in the simulated HCS crossing. Notably, considering the very high orbital speed of PSP, a timing mismatch of even a single day can result in significant discrepancies due to the considerable radial distance covered by the spacecraft within that period.

The measurements of $V_R$ indicate that PSP observed three HSSs back-to-back during the inbound phase. Although there was a data gap from January 7 to 8, 2020, the structure before and after this gap suggests a smooth rising and trailing edge of the HSS. Additionally, the solar wind speed near the perihelion dropped below $300\, \mathrm{km}\,\mathrm{s}^{-1}$ and remained relatively stable until the arrival of the next HSS on January 29, 2021. On the other hand, aside from accurately reconstructing the arrival and leading edge of the first HSS, the simulation did not resolve the individual fast streams during the inbound phase. Instead, it depicted a single continuous HSS with a long trailing edge extending until the perihelion. However, the simulation successfully reproduced the general trend in the slow wind interval after the perihelion, though it overestimated the speed by about $50 \, \mathrm{km}\,\mathrm{s}^{-1}$, a behavior consistent with previous PSP perihelia. Furthermore, the ensemble simulation accurately reconstructed the HSS arrival time on January 29, 2021, along with its structure, within the range of uncertainties. It is important to note that the level of uncertainty varied throughout the period, with relatively larger uncertainties around $200\, \mathrm{km}\,\mathrm{s}^{-1}$ in speed observed for the HSSs on January 5 and January 31, 2020.

PSP observed a typical proton density profile with a gradual increase during the inbound phase to perihelion and a decline during the outbound phase, reflecting its dependence on heliocentric distance. On perihelion day, the density enhancement at the HCS crossing, combined with the closer heliocentric distance, resulted in a peak of about $3.2 \times \mathrm{10^{3}} \, \mathrm{cm^{-3}}$. Conversely, the ensemble simulation captured the density distribution during the tail ends and gradual leading edge, but consistently underestimated values elsewhere. A timing mismatch in HCS crossings on perihelion day caused the discrepancy of about $1.6 \times \mathrm{10^{3}} \, \mathrm{cm^{-3}}$, while the discrepancy of more than $\mathrm{10^{3}} \, \mathrm{cm^{-3}}$ around January 19, 2020, resulted from the incorrectly simulated unobserved fast wind with lower density.

Furthermore, the ensemble simulation captured the average trend, but did not fully reproduce the variations in proton temperature associated with the HSSs during January 9–13, 2020.  Otherwise, the distributions were successfully reconstructed in the inbound phase. It reproduced the distributions near perihelion and the outbound phase reasonably well within the range of uncertainties. Note that temperature and density measurements for the HSS from January 29 to February 2, 2021, were not shown due to poor data quality.

Figure~\ref{fig:P7}(b) shows that Earth experienced predominant negative polarity during P7. Frequent oscillations of $B_R$ between positive and negative values, along with multiple IMF polarity reversals throughout the period, indicate that the Earth traversed the SW close to the HCS. Moreover, the polarity reversals on January 11, 15, 17, 18, 22, 25, and February 01, 2021, coincide with the SB crossings, indicating that the spacecraft observed a multiple-sector structure. Additionally, the distribution of $B$ shows four distinct field enhancements around January 06, 11, 26, and February 02, 2021, with magnitudes of $\sim \! \! 10 \mathrm{nT}$, caused by the compression regions instigated by the interaction of the HSS with the slow wind. In comparison, the ensemble simulation reproduced the overall profile of $B$ and captured three out of four compression regions reasonably well, with a slight timing mismatch for the third one. However, while the simulation reconstructed the predominant negative polarity, it did not capture most of the HCS crossings, except for those between January 11 and 15, 2021, and on February 01, 2021.

Furthermore, the plasma parameters indicate that three HSSs of distinct origins, centered around January 7, January 12, and January 27, 2021, were observed at Earth. The first and third HSSs, with negative polarity, originated from the southern hemisphere, while the second, with positive polarity, came from the northern hemisphere. A sharp increase in plasma and magnetic field variables on the final day of the period indicates another HSS, although only its compression region was visible.

In comparison, the simulation captured the first and third HSSs reasonably well, but did not capture the second. The simulation accurately reconstructed the sharp rise, amplitude, and timing of the first HSS, but overestimated the trailing edge by $\sim \! \! 50\, \mathrm{km}\,\mathrm{s}^{-1}$. The simulation indicated an early arrival of the third HSS by more than half a day and did not reproduce the variations and peak magnitude while systematically overestimating the trailing edge. The second observed HSS was completely missed, resulting in a discrepancy of approximately $150 \, \mathrm{km}\,\mathrm{s}^{-1}$. A discrepancy of similar magnitude occurred between January 16 and 19, 2021, due to the presence of an unobserved faster stream in the simulation.

The simulation reasonably reconstructed the overall profiles of proton density and temperature. However, the missed second HSS and its associated compression regions during January 11–16, 2021, led to the largest discrepancies, with errors of approximately $40\, \mathrm{cm^{-3}}$ in proton density and $0.3\,\mathrm{MK}$ in temperature. Although the simulation reproduced $V_R$ for the third HSS relatively well, it did not capture the corresponding temperature and density profiles.

Figure~\ref{fig:P7}(c) shows that STEREO-A observed a dominant positive polarity, consistent with its trajectory above the heliographic equator during P7. The spacecraft experienced multiple polarity reversals as it traveled near the HCS, with reversals on January 7, 20, 26, and 29, 2021, coinciding with SB crossings, highlighting the multi-sector structure. In comparison, the ensemble simulation successfully reproduced $B$ and $|B_R|$s, their variations, and IMF polarity sectors, capturing all SB crossings reasonably well within uncertainty bounds. Notably, the ensemble spread in $B$ and $|B_R|$ increased near the SB crossings, along with a corresponding increase in the temporal spread of the IMF polarity sector boundaries.

Plasma observations indicate that STEREO-A predominantly observed slow SW during the first half of P7, followed by two HSSs, centered around January 20 and 31, 2021, during the second half. However, the ensemble simulation did not reproduce the observed slow wind and instead produced multiple distinct fast streams, leading to a discrepancy of $\sim \! \! 100 \, \mathrm{km}\,\mathrm{s}^{-1}$ from January 8–19, 2021. Additionally, the simulation accurately captured the arrival of the first HSS, but not its amplitude, while successfully reconstructing its trailing edge. The second HSS was captured, but its leading edge and peak were underestimated.

A comparison of proton density shows that while the ensemble simulation accurately captured the general profile, it overestimated the amplitude of the compression region centered on January 6, 2021, by about $20\, \mathrm{cm^{-3}}$. Finally, the proton temperature for the first half of the period was reproduced reasonably well by the simulation, despite overestimating the speed. Discrepancies of approximately $0.16\, \mathrm{MK}$ and $0.08\, \mathrm{MK}$ occurred around January 20 and 31, 2021, in the second half of the period due to the underestimation of HSS temperatures.

From Figure~\ref{fig:P7}(d), it is evident that SolO spent nearly equal time in both IMF polarity sectors and experienced several polarity reversals in the second half of P7, indicating its proximity to the HCS. During this period, the spacecraft encountered SB crossings on January 5, 11, 14, 25, and 30, 2021. Notably, due to the unavailability of magnetic field data from January 17 to 21, 2021, the exact timing of the SB crossing from negative to positive polarity cannot be determined. 

In comparison, the distributions of $B$ and $|B_R|$ were well reproduced by the ensemble simulation within the range of uncertainties. However, around January 12, 2021, a discrepancy of approximately $14 \, \mathrm{nT}$ in $B_R$ occurred due to a $1.5$-day delay in the simulated SB crossing. Nevertheless, apart from a few delayed HCS crossings, the simulation accurately captured most SB crossings within the uncertainty bounds. Unfortunately, SW plasma data were unavailable during P7, preventing a direct evaluation of the simulation. However, the ensemble spread in plasma variables indicates that uncertainty was relatively higher between January 20–26, 2021, than for other periods.

\subsubsection{Perihelion Pass 8 (CRs 2243-2244)}

\begin{figure*}[ht!]
\centering
\includegraphics[width=\textwidth]{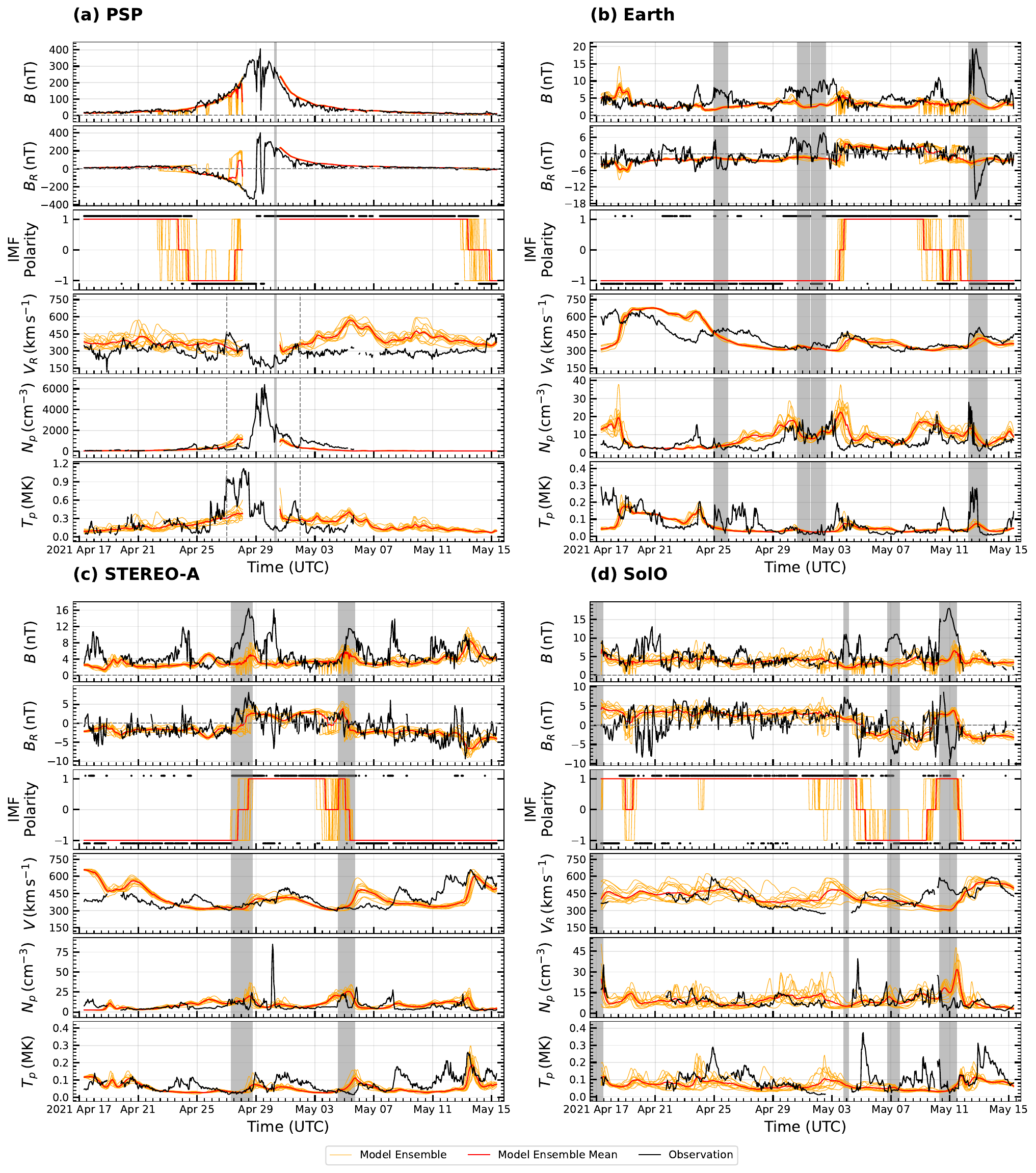}
\caption{Comparison of our ensemble simulations with in situ observations along the trajectories of PSP (a), Earth (b), STEREO-A (c), and SolO (d) during P8, spanning from April 17 to May 15, 2021. The same format as in Figure \ref{fig:P4}.}
\label{fig:P8}
\end{figure*}

As shown in Figure~\ref{fig:P8}(a), PSP was predominantly in a positive IMF polarity environment during P8, except for brief periods of negative polarity during the five days before perihelion and the final two days of the outbound phase. Compared to previous perihelia, PSP observed double HCS crossings that occurred within a short time frame marked by sharp transitions from negative to positive polarity and back. Notably, the amplitude of \( B_R \) reached \(\pm 400 \, \mathrm{nT}\), exceeding the values observed in the previous perihelia, as PSP approached the Sun as close as \( 15\, \mathrm{R_\odot} \) during perihelion. This marked the first in situ measurement inside the sub-Alfv\'enic region. 

In comparison, the simulation of the IMF polarity and its radial component aligns well with observations, remaining within the range of uncertainties. Our ensemble simulation successfully reconstructed the SB crossings on April 25 and May 14, 2021, while producing an early HCS crossing, approximately a day before the one observed near the perihelion. However, since our IHS simulation boundary starts at \( 21.5\, \mathrm{R_\odot} \), no solution exists for a brief period near perihelion, producing a gap in the comparison plots.

PSP measurements of \( V_R \) indicate that the spacecraft was immersed into a slow SW environment, with slight variations during the inbound phase, reaching speeds below \(200 \, \mathrm{km}\,\mathrm{s}^{-1}\) near perihelion. No significant deviations from an average speed of \(300 \, \mathrm{km}\,\mathrm{s}^{-1}\) were observed during the outbound phase, except for the final day when an HSS arrived. Additionally, the proton density peaked at approximately \(6 \times 10^3 \, \mathrm{cm^{-3}}\) on the day of perihelion before returning to typical levels for the remainder of the period. Finally, the SW temperature closely followed the velocity distribution throughout the entire period.

In contrast, the simulation showed a good agreement with the plasma observations during the inbound phase while overestimating the radial speed by approximately \(150\, \mathrm{km}\,\mathrm{s}^{-1}\) due to two unobserved HSSs centered on May 5 and 10, 2021, during the outbound phase. Furthermore, the proton density was underestimated by approximately \(1000 \, \mathrm{cm^{-3}}\), while their temperature was slightly overestimated for the first unobserved HSS. Unfortunately, a gap exists in the density and temperature observations after May 6, 2021, due to the data quality issues.

Notably, the ensemble spread for \( V_R \) and \( T_P \) was approximately the same both during the inbound and outbound phases, while the spread of density was larger in the first half of P8 than in its second half. Finally, the IMF exhibited the largest spread near the HCS/SB crossings compared to the rest of the period.

Figure~\ref{fig:P8}(b) illustrates that Earth predominantly traveled through the negative IMF polarity region during the first half of P8 before transitioning to the positive polarity in the second half. The distribution of $B_R$, fluctuating near zero, suggests that Earth was grazing the HCS. Notably, four distinct ICMEs were observed at Earth, indicating increased solar activity, which primarily constrained our validation to the non-ICME portions of the perihelion pass. In comparison, the SB crossings observed on May 3, 10, 11, and 12, 2021, were reasonably well reproduced, whereas the crossings on April 21 and 29, 2021, were not captured by the ensemble simulation. Although the ensemble simulation did not fully capture the variations in $B_R$ during the first half of the period, it successfully reproduced the overall trend and better reproduced the fluctuations in the second half.

Furthermore, the plasma parameters indicate the presence of an HSS centered on April 19, 2021, with a peak speed exceeding $600\, \mathrm{km}\,\mathrm{s}^{-1}$, along with two additional HSSs centered on April 26 and May 4, 2021, exhibiting broad peak speeds of approximately $500\,\mathrm{km}\,\mathrm{s}^{-1}$. The IMF polarities suggest that the first two HSSs originated from the southern hemisphere, while the third one originated from the northern hemisphere. Due to the selected validation window, only the central portion and trailing edge of the first observed HSS are visible in the figure. The proton density and temperature profiles indicate compression regions associated with the latter two HSSs, which appear as enhancements at their leading edges.

In contrast, the ensemble simulation reasonably reproduced $V_R$ during the second half of the period and accurately captured the amplitude and trailing edge of the HSS centered on April 26, 2021. However, the arrival time of this simulated HSS was delayed by one day compared to the observations, this discrepancy is also evident in the IMF polarity. During the first half of the period, the simulation did not capture the first two observed HSSs. Instead, a single broad HSS with a peak speed of $600\, \mathrm{km}\,\mathrm{s}^{-1}$ was generated, resulting in discrepancies of approximately $200\, \mathrm{km}\,\mathrm{s}^{-1}$ in $V_R$, $10\, \mathrm{cm}^{-3}$ in $N_p$, and $0.05\, \mathrm{MK}$ in $T_p$.

Figure~\ref{fig:P8}(c) shows that STEREO-A predominantly traversed the negative polarity sector, as it remained below the heliographic equator throughout the period. It experienced SB crossings from negative to positive polarity on April 27 and vice versa on May 6, 2021, resulting in a three-sector polarity structure. Notably, both SB crossings coincided with the arrival of ICME. These crossings were due to the HSS centered on April 30, 2021, which carried positive magnetic field polarity as it originated from the northern hemisphere. The IMF polarity and speed suggest that the same HSS arrived at Earth after 4–5 days due to co-rotation. Additionally, STEREO-A observed multiple other HSSs centered on April 20, 25, May 9, and 13, 2021, with negative polarities as they originated from the southern hemisphere. The distributions of $B$, $N_p$, and $T_p$ show multiple enhancements centered around April 18, 24, and 30, May 9, 11, and 13, 2021, across P8, primarily due to the compression regions formed due to the interactions of the slow and fast SW streams. However, enhancements in $N_p$ are not as pronounced relative to $B$ and $T_p$ due to a density peak of $\sim \! \! 85\, \mathrm{cm^{-3}}$ that appeared on May 1, 2021, coinciding with the transient HCS crossing.

On the other hand, our ensemble simulation successfully reproduced the IMF polarity and $B_R$ by capturing both SB crossings. However, only the overall profile of $B$ was reconstructed, while the field enhancements associated with the compression regions were not captured, which resulted in a discrepancy of approximately $\sim 8\, \mathrm{nT}$. The SW speed between the two SB crossings was also well reproduced, as the simulation captured the HSS responsible for it. In contrast, the simulation completely missed the HSSs that occurred between April 23–27 and May 8–13, 2021, which resulted in discrepancies of about $200\, \mathrm{km\,s^{-1}}$. Additionally, it produced an unobserved HSS around May 6, 2021, and a double-peaked profile for the first HSS, with mismatched leading and trailing edges. The overall trend in proton density was reproduced, although the peak on April 30, 2021, was not captured due to the missed transient HCS crossing. Finally, the simulation successfully reproduced the enhanced proton temperature profile during the first and last HSSs but did not capture it for the other missed HSSs.

As shown in Figure~\ref{fig:P8}(d), SolO primarily traversed into the positive magnetic polarity sector during P8, accompanied by multiple polarity reversals. The polarity reversals observed on April 18, 19, and 21, as well as May 4, 9, and 11, 2021, correspond to SB crossings, indicating a multi-sector IMF environment. Additionally, the spacecraft encountered ICMEs on April 17, May 4, 7, and 11, 2021. The enhancement in the $V_R$ and $T_p$ suggests that SolO encountered HSSs centered on April 25, May 5, and 14, 2021. Unfortunately, data gaps exist in the magnetic field measurements from May 12-13, 2021, and in plasma parameters from mid-April 17-22 and May 2-4, 2021.

In comparison, our ensemble simulation reasonably reproduced the distributions of $B$ and $B_R$ within the range of uncertainties, except during April 17-22, 2021, where the compression region and polarities were not accurately simulated. However, all SB crossings were captured by the simulation, albeit with slight timing mismatches. The simulation successfully reconstructed the overall profile of $V_R$ by capturing the first and final HSSs but missed the intermediate ones. The density was also well reconstructed, except for the sharp peak on May 5, 2024, where the leading edge of the HSS was missed, as indicated by $V_R$. Finally, the simulation missed most of the temperature enhancements associated with HSSs.

\subsubsection{Perihelion Pass 9 (CRs 2246-2247)}

\begin{figure*}[ht!]
\centering
\includegraphics[width=\textwidth]{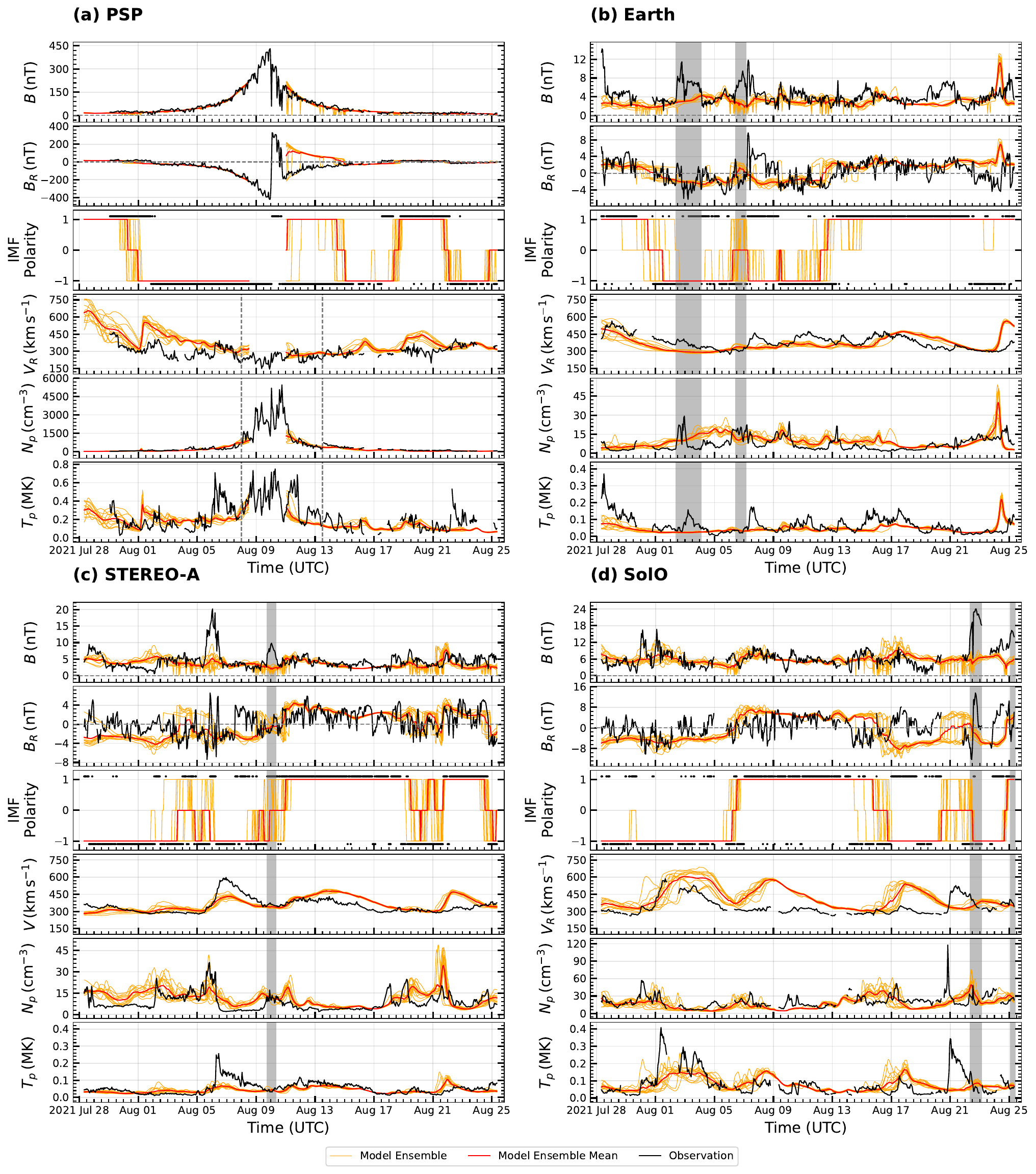}
\caption{Comparison of our ensemble simulations with in situ data along the trajectories of PSP (a), Earth (b), STEREO-A (c), and SolO (d) during P9, from July 28 to August 25, 2021. The same format as in Figure \ref{fig:P4}.}
\label{fig:P9}
\end{figure*}

Figure~\ref{fig:P9}(a) shows that PSP observed $B$ and $B_R$ following a trend similar to the previous perihelion passes, with a slightly increased amplitude, as $B$ reached approximately $420\, \mathrm{nT}$ on the day of perihelion. During P9, PSP reached heliocentric distance of $15\, \mathrm{R_\odot}$, the same as for P8. PSP predominantly traversed a negative IMF polarity environment, with brief excursions into the positive polarity sector. Polarity reversals on July 30, August 10, 17, and 22, 2021, corresponded to SB crossings by the spacecraft. Notably, on August 10, 2021, two consecutive polarity reversals occurred within a single day, which, due to PSP’s high orbital velocity, still coincided with SB crossings.

In comparison, the distributions of $B_R$ and $B$ were accurately reproduced by the simulation during the inbound phase. However, during the outbound phase, a large discrepancy in $B_R$ was observed between the ensemble mean of the simulation and the observations from August 11–15, 2021, with a maximum error of approximately $200\, \mathrm{nT}$ due to a mismatch in IMF polarity. Nevertheless, the simulation successfully reproduced the overall IMF polarity sector structure for the remainder of the period, with a slight timing mismatch in the HCS/SB crossings. Specifically, the SB crossing on August 1, 2021, occurred earlier, while the one on August 17, 2021, was delayed by one day in the simulation compared to the observations. Additionally, the simulation results near PSP’s perihelion are not shown in the figure, as PSP was located below the inner boundary of the simulation.

Plasma parameters from PSP indicate that the spacecraft observed a steady, SW with an average speed of $300\, \mathrm{km} \,\mathrm{s^{-1}}$ throughout P9, along with the two instances of HSSs exhibiting relatively small speed enhancements centered around July 30 and August 2, 2021. The proton density peaked at approximately $5.5 \times 10^{3} \,\mathrm{cm^{-3}}$ near perihelion, similar to the previous perihelion pass.

On the other hand, the ensemble simulation reasonably reproduced the slow SW and the trailing edge of the first HSS within the range of uncertainties, while overestimating both the peak speed and trailing edge of the second HSS. Proton density and temperature were also generally well reproduced. There are, however, some notable discrepancies of approximately $0.25\, \mathrm{MK}$ around August 6, 2021, and August 22, 2021, despite the accurate reproduction of $V_R$ distributions during these periods.

Figure~\ref{fig:P9}(b) shows that during P9, Earth was predominantly immersed in the positive polarity sector, as it remained above the heliographic equator throughout this period, as shown in Figure~\ref{fig:MHD_2D_slices}. Observations of SB crossings on July 31, August 3, 5, 7, 9, 13, 22, and 25, 2021, indicate that Earth traversed a multi-sector SW environment. Notably, two ICMEs coincided with the SB crossings on August 3 and 7, 2021. Multiple peaks in $V_R$  with relatively moderate speed enhancements of $\sim 150\, \mathrm{km\,s^{-1}}$, centered around August 7, 11, and 17, were observed, along with the relatively faster SW  at the start of the period.

In comparison, our ensemble simulation reproduced the general profile of $B_R$ except around August 7 and 23, 2021, where mismatches in the simulated IMF polarities occurred due to missed SB crossings. Additionally, the simulation reproduced only some of the enhancements in $V_R$ and $T_p$. For example, the simulation completely missed the first two moderate peaks, while the last one was captured with a slight delay. Although the simulation reproduced the general trend and variability in proton density, it overestimated the overall profile. 

As shown in Figure~\ref{fig:P9}(c), STEREO-A grazed through the HCS during P8 while predominantly observing a negative IMF polarity. SB crossings on August 9, 19, 21, and 24, 2021, indicate that the spacecraft encountered a multi-sector SW environment. Plasma variables suggest that a predominantly slow SW was observed, accompanied by an HSS with a gradient in the speed of $\sim 300\, \mathrm{km \, s^{-1}}$ and temperature of about $0.2\, \mathrm{MK}$ during August 5–11, 2021. Consequently, compression regions caused by the HSS are evident in the distribution of $B$ and $N_p$, appearing as enhancements at the leading edge.

On the other hand, the ensemble simulation successfully captured the IMF polarity by reproducing all SB crossings within the range of uncertainties. The distribution of $B$ was well reproduced for most of the period, except for the compression region around August 5, 2021, where the field enhancement was not captured. Additionally, the general profile and variations of $B_R$ were reconstructed reasonably well during the second half of the validation period but were slightly overestimated in the first half. The simulation also reproduced a predominantly slow SW during the first half but underestimated both the peak and gradient of the HSS speed and temperature. In the second half of P9, two unobserved HSSs with moderate speed gradients were obtained instead of the slow SW. The overall trend in proton density was reproduced well within the range of uncertainties, although it was overestimated during the HSS.

Figure~\ref{fig:P9}(d) shows that SolO spent nearly equal time in both polarity sectors during P9. It experienced several transient polarity reversals throughout the period, along with SB crossings on August 7, 14, 17, 19, and 24, 2021. The spacecraft was predominantly immersed in slow SW while encountering two HSSs centered around August 3 and August 21–23, 2021. The former spanned longer with a smooth gradient than the latter, although both originated in the southern hemisphere. Compression regions formed at the slow and fast stream interfaces are evident in the density enhancements. The amplitude of the latter compression region was substantially higher than the former one due to the sharper speed gradient. There is a data gap in the magnetic field for most of the second HSS, limiting a complete analysis. However, field enhancements in the first compression region are still visible in the distribution of $B$.

In comparison, the ensemble simulation reasonably reproduced the IMF distribution within the range of uncertainties, but did not accurately reconstruct $B_R$. A discrepancy of approximately $16, \mathrm{nT}$ in $B_R$ between August 13–24, 2021, primarily resulted from the mismatches in the simulated IMF polarity due to missed SB crossings. However, the simulation captured other SB crossings with a slight timing mismatch. Moreover, the ensemble simulation reproduced only the first HSS within the range of uncertainties but missed the second one, leading to large discrepancies of approximately $\sim\! \! 200\, \mathrm{km\,s^{-1}}$ in $V_R$ and $\sim\! \! 0.3\, \mathrm{MK}$ in $T_P$. Consequently, the sharp density peak of $120\, \mathrm{cm^{-3}}$ at the second stream interface was also not captured by the simulation. Additionally, the simulation produced two unobserved HSSs centered on August 9 and 18, 2021, during the predominant slow SW phase, resulting in a speed discrepancy of about $300\, \mathrm{km\,s^{-1}}$.

\section{Discussion and Conclusion}\label{sec:DandC}

Here we presented 3D MHD simulations of the supersonic ambient SW in the IHS using time-dependent inner boundary conditions derived from the WSA coronal model. The WSA model was driven by synchronic photospheric maps generated by the ADAPT model, which was constrained by HMI LOS magnetograms, ensuring a more realistic representation of magnetic fields on the solar surface. To quantify the effect of the inherent uncertainty in the photospheric boundary conditions on SW simulations, we employed an ensemble modeling technique and performed multiple runs using different realizations provided by ADAPT. The latter were generated by perturbing the distributions of supergranular flows, particularly on the far-side and polar regions of the photosphere.

To evaluate the performance of the ADAPT-WSA-MS-FLUKSS model, numerical simulations were compared with in situ observations along the PSP, Earth, STEREO-A, and SolO trajectories in the IHS during the rising phase of Solar Cycle 25 (2020–2021). The key variables, i.e., magnetic field strength, the radial component of the magnetic field vector, IMF polarity, SW radial velocity component, and plasma number density and temperature, were used for a time series analysis. We analyzed six different time intervals, P4 to P9, each covering approximately one solar rotation around the fourth to ninth PSP perihelia, so our validation took advantage of the diverse radial, longitudinal, and latitudinal configurations of multiple spacecraft. 

We found that the simulation reproduced the overall profile of the magnetic field more accurately than the plasma variables for all spacecraft and perihelion passes. This is likely because magnetic field measurements, which are obtained with reasonable accuracy in the photosphere through remote sensing, are used to constrain the flux transport model (ADAPT) and coronal model (WSA). These results outline the effectiveness and limitations of the potential field models, PFSS and PFCS, in capturing the coronal magnetic field structures during the near-minimum phase of the solar cycle. Furthermore, our analysis shows that the simulations reproduced the overall profile of $B$ more accurately than that of $B_R$. Most discrepancies in $B_R$ were due to the polarity mismatches rather than the errors in the magnitude of $B_R$. This is because $|B_R|$ was also scaled when implementing our inner boundary conditions for the ``open flux'' problem. The scaling of $|B_R|$ by a constant factor (2 in our case) was intended to help the model match the Earth-based magnetic field observations. It also performed well at STEREO-A and SolO for all perihelion passes. However, at PSP, from P4 to P8, the model slightly overestimated $B$ and $|B_R|$ near perihelia and during the outbound phases, whereas a good agreement with observation during P9. These findings suggest that ad hoc scaling by a constant factor may sometimes overestimate the field strength very close to the Sun. Interestingly, despite SolO not always being near a $1\,\mathrm{au}$ orbit, the scaling of $|B_R|$ still worked well for its observations.

Accurate modeling of the IMF polarity is essential for reconstructing the plasma and magnetic field properties of the ambient SW at any spacecraft. To evaluate the ADAPT-WSA-MS-FLUKSS model performance in reproducing IMF polarity, we analyzed its ability to capture large-scale HCS crossings, particularly those coinciding with SB crossings. Our detailed analysis indicates that the model captured most events reasonably well for different spacecraft and perihelion passes. However, a few crossings were missed. Notably, the model performance at PSP improved after P6 compared to the earlier passes,  a similar trend being observed at STEREO-A. In contrast, the simulation consistently reproduced SB crossings at SolO for most time intervals. At Earth, however, the reconstruction of SB crossings varied for different perihelion passes. Considering all perihelion passes, the best performance was observed at STEREO-A, likely because the connecting IMF field lines originated from the Sun’s near side, where magnetic field observations are continuously available. Additionally, improvements in the model performance after P6 are likely associated with the gradually increasing HCS tilt at the inner boundary of the heliospheric model, which was initially very flat, leading to more significant variability in the SW at spacecraft latitudes as the solar cycle progressed from P4 to P9.

Further, our ensemble simulations at multiple spacecraft showed that the spread among ensemble members increases near SB crossings, as reflected in the IMF polarity panels. This aligns with the fact that the largest variation in $B_R$ is concentrated near the HCS, as illustrated in Figure \ref{fig:bcstd}, which presents the standard deviation maps of $B_R$ at the inner boundary. Across all spacecraft and perihelion passes, the temporal spread near SB crossings was typically about $1$ to $1.5$ days, which agrees well with the typical SB crossing time error of the ensemble mean. This highlights the significant influence of supergranular distributions in the polar and far-side regions on the solar surface within the flux transport model in shaping the HCS and, consequently, the coronal magnetic field. Notably, even when the ensemble mean missed certain SB crossings, individual ensemble members often succeeded in capturing them. These results underscore the importance of uncertainty quantification and demonstrate its advantage over purely deterministic modeling approaches.

Regarding plasma variables, the accuracy of reconstructing SW radial velocity is expected to correlate with plasma density and temperature, since both are derived at the inner boundary of the MHD model using the conservation of momentum flux and thermal pressure balance. As anticipated, our results show that when the simulated radial velocity component is overestimated, the plasma density and temperature are underestimated and overestimated, respectively. Among these three variables, temperature was the least reliably reconstructed. For example, during January 25–29, 2021 (P7), and July 28–30, 2021 (P9) at Earth, the simulation reasonably captured the radial velocity and density, while the temperature profile associated with fast SW was not well represented. Notably, our results are in line with \cite{Jianetal:2015}, which highlights the discrepancies in reconstructing the temperature distributions in most of the models implemented at CCMC. This suggests that further model refinement is needed to estimate the plasma temperature at the inner heliospheric boundary, possibly by improving the parameterization used to derive it from the radial velocity and density.

In addition, our simulation, performed for multiple perihelion passes, shows that the radial velocity component of the predominantly slow SW observed at PSP are systematically overestimated in contrast to the other locations in the IHS. To compensate for the unrealistic acceleration of SW in the WSA model, we applied a uniform reduction of $75\, \mathrm{km\,s^{-1}}$ to the asymptotic radial velocity at the inner boundary. However, our results indicate that this ad hoc correction is sufficient to match the MHD simulations near the $1,\mathrm{au}$, but not at PSP. It is important to note that the current version of the WSA model was originally optimized for near-Earth observations. A recent study by \cite{Samaraetal:2024} similarly found that calibrating the WSA speed for PSP does not yield accurate results at Earth, and vice versa. These findings highlight the importance of multi-point validation and the limitations of applying a single calibration across different heliospheric locations.

Furthermore, the ensemble simulation captured most observed HSSs reasonably well, with only minor differences in arrival time, amplitude, and structure. Performance varied across spacecraft and perihelion passes without any systematic dependence on longitudes. Notably, the simulation produced more unobserved HSSs than it missed observed ones. These unobserved cases can be broadly grouped based on whether the IMF polarity was consistent with observations or not. In some instances, such as P7 at STEREO-A (January 13–20, 2021), P8 at PSP (May 3–14, 2021), and P9 at SolO (August 6–13, 2021), the IMF polarity was correctly reproduced, yet unobserved HSSs were generated. In contrast, in cases like P5 at Earth (May 27–30, 2020) and P9 at SolO (August 18–21, 2021), the generation of unobserved HSSs coincided with incorrect IMF polarity. These findings point to certain biases in the determination of SW velocity distribution at the inner boundary, which contributes to the generation of unobserved HSSs. For example, WSA sometimes underestimates the latitudinal extent of the slow SW, as illustrated in Figure \ref{fig:bc}.
 
Overall, our comprehensive analysis of uncertainties associated with the applied model, provided a robust assessment of the model performance across diverse radial, longitudinal, and latitudinal configurations during the rising phase of Solar Cycle 25. Additionally, it offered critical insights into the impact of uncertainties in the photospheric boundary conditions, specifically those arising from perturbations in the supergranular distribution in far-side and polar regions, across various locations in the IHS.

In our ensemble analysis, only uncertainties associated with the ADAPT model were taken into account. However, the WSA model itself and the input magnetograms are also prone to such uncertainties. The WSA model typically assumes deterministic values for the key parameters, which are, in fact, uncertain and could cause a large spread in the SW speed \citep[e.g.,][]{Issanetal:2023}. There are also considerable differences between magnetograms from different sources \citep{riley2013, Rileyetal:2014}. The role of those uncertainties should be investigated separately in the future.

We have demonstrated that it is practically impossible to reproduce all observations, even at each location, not to mention doing this consistently at multiple spacecraft and at Earth. The reasons for that should be associated with the availability of quality boundary conditions. Shortly formulated, the entire solution for the SW properties and IMF in the IHS is fully determined by the time-dependent inner boundary conditions. One source of uncertainty is clearly associated with the photospheric magnetic field data provided by ADAPT. The corresponding ensemble members create a spread in the distributions, but it is frequently insufficient to embrace the complicated distributions available from observational data. Another source of uncertainties is due to the assumptions made in the coronal model employed to determine the boundary conditions for our simulations. The WSA model is essentially a kinematic data-assimilation model, which requires ad hoc corrections to make the boundary conditions consistent with the further MHD simulation in the IHS. While it is reasonable to expect that more sophisticated models will ultimately improve the situation,  the uncertainties in the observational input provided by the SFT models will not disappear. It is of importance, however, to evaluate the capabilities of ensemble simulations and potentially combine them with machine learning techniques to improve space weather forecasts, as this is done, e.g., in modeling CME arrival times in \citet{Singhetal:2022}. 

Ensemble modeling of the ambient SW to characterize the inherent uncertainties in the boundary conditions is critically advantageous for advancing the physical understanding of the SW propagation from the Sun to Earth, which is characterized by a number of transient phenomena. First, it allows one to capture intrinsic uncertainties in the synchronic photospheric magnetic maps, which are inevitable because of the absence of full-Sun data. These uncertainties propagate non-uniformly through the heliosphere, and our results demonstrate that the ensemble spread varies significantly with heliospheric location. This spatial variability is particularly important for understanding CME–SW interactions, which are often unaccounted for in models that rely on a single, deterministic SW background. Furthermore, ensemble simulations reveal significant variations in the time of arrival for such key structures as the HCS and HSSs, both of which are known to strongly influence CME dynamics, at different spacecraft and planets. By performing ensemble modeling, we ensure that such interactions are neither missed nor misrepresented. This approach also makes it possible to perform a deeper investigation of how the supergranular flow within the SFT model can influence the background SW solutions. Until full-solar-surface magnetic field measurements become available through future missions (e.g., those proposed in the NASA Heliophysics Decadal Survey 2025 \cite{Decadal_Survey:2024}), ensemble modeling remains essential for generating a physically realistic and interpretable view of the heliospheric environment.

From a forecasting perspective, this ensemble modeling approach significantly improves our ability to assess and communicate uncertainties in the ambient SW. By simulating a range of SW realizations, which are all consistent with the current state of the photospheric magnetic field, one can quantify the confidence bounds for any specific prediction. This is particularly important because relying on a single, even the best performing, ensemble member can result in a forecast that omits the key structures and interactions, e.g., CME compression by a preceding HSS or its deflection near the HCS. These features can affect CME arrival times and their geo-effectiveness. Our study represents the first step towards understanding of how uncertainties in the input synchronic maps propagate through the model chain. Importantly, such modeling also helps distinguish between the features that are robust across the ensemble members and those that are sensitive to boundary uncertainties, enabling forecasters to better assess the associated risks. As demonstrated in the CME propagation analysis of \citet{Singhetal:2025a}, uncertainties attributable to different components of the modeling chain, can cancel out or add up, ultimately affecting the final prediction. Thus, ensemble simulations are not only valuable for probabilistic forecasting but are also necessary for building more complete and resilient SWx prediction systems. Here, we analyzed the discrepancies between simulations and observations only qualitatively. A more detailed quantitative analysis will be presented in the followup paper.

\begin{acknowledgments}

The authors acknowledge support from the joint NSF-NASA Space Weather with Quantified Uncertainties (SWQU) Program through  NSF award AGS-2028154 and NASA grant 80NSSC20K1582. We also acknowledge support from the PSP mission through the UAH–SAO agreement SV4-84017 and NASA R2O2R grant 80NSSC22K0270. DVH is grateful for the support provided by the NASA FINESST grant 80NSSC22K0058. TKK and NVP acknowledge additional support from the NASA Grant 80NSSC20K1453. CNA was supported in part by the NASA SWQU grant listed above and the NASA competed Heliophysics Internal Scientist Funding Model (ISFM). Supercomputing resources were made available to us by the NASA High-End Computing Program awards SMD-20-92772410 and SMD-21-44038581, as well as the ACCESS project MCA07S033. We acknowledge using spacecraft trajectory data from \url{https://omniweb.gsfc.nasa.gov/coho/helios/heli.html}.  The authors acknowledge using SDO/HMI data from the Joint Science Operations Center (\url{http://jsoc.stanford.edu/}). We also acknowledge the use of OMNI, STEREO-A, and PSP/SPAN-I data from NASA’s Space Physics Data Facility (SPDF) CDAWeb services (\url{https://cdaweb.gsfc.nasa.gov/}). Additionally, we have utilized PSP/SPC and PSP/FIELDS data, which were obtained from the following link:\url{https://spdf.gsfc.nasa.gov/pub/data/psp/}. We also wish to acknowledge the use of SolO data from the Solar Orbiter Archive (\url{https://soar.esac.esa.int/soar/}). Furthermore, this work has benefited from data produced through the collaboration between the Air Force Research Laboratory (AFRL) and the National Solar Observatory (NSO). Specifically, the ADAPT maps employed in our study were accessed from \url{https://gong.nso.edu/adapt/maps/special/psp/adapt_hmi-los/}. We also acknowledge using the HELIO4CAST ICME catalog \citep{Moestl:2020}.

\end{acknowledgments}

\clearpage

\bibliography{hegde}{}
\bibliographystyle{aasjournal}

\end{document}